\documentclass[twocolumn,10pt,letterpaper]{article}

\usepackage[letterpaper,margin=1.5cm,footskip=0.85cm]{geometry}
\usepackage{graphicx}
\usepackage{amsmath,amssymb,amsfonts}
\usepackage{mathrsfs}
\usepackage{stmaryrd}
\usepackage{booktabs}
\usepackage{xcolor}
\usepackage{textcomp}
\usepackage{microtype}
\usepackage[hang,flushmargin]{footmisc}
\usepackage{authblk}
\usepackage{titlesec}
\titleformat{\section}[hang]
  {\normalfont\large\bfseries}{\thesection}{0.7em}{\setlength{\parindent}{0pt}}
\titleformat{\subsection}
  {\normalfont\normalsize\bfseries}{\thesubsection}{0.75em}{}
\titleformat{\subsubsection}
  {\normalfont\normalsize\itshape}{\thesubsubsection}{0.75em}{}
\titlespacing*{\section}{0pt}{4ex plus .5ex}{1ex}
\titlespacing*{\subsection}{0pt}{1.5ex plus .5ex}{0.75ex}

\usepackage[font=small,labelfont=sc,labelsep=period]{caption}
\usepackage{etoolbox}

\makeatletter
\patchcmd{\@startsection}
  {\@afterindenttrue}
  {\@afterindentfalse}
  {}{}
\makeatother

\apptocmd{\thebibliography}{%
  \setlength{\itemsep}{0pt}%
  \setlength{\parskip}{1pt}%
}{}{}
\let\oldthebibliography\thebibliography
\renewcommand{\thebibliography}[1]{%
  \oldthebibliography{#1}%
  \small
}

\usepackage[switch]{lineno}

\usepackage{hyperref}
\newcommand{\meth}[1]{Methods, Sec.~\ref{#1}}
\date{}

\title{From the possible to the optimal: differentiable optimization of gene regulatory networks}

\author[1]{Leah Friedman}
\author[2]{Ga\v{s}per Tka\v{c}ik}
\author[1,3]{Thomas Gregor}
\author[4,5,6]{Paul Fran\c{c}ois$^{*}$}

\affil[1]{Department of Developmental and Stem Cell Biology,\protect\\
CNRS UMR3738 Paris Cit\'e, Institut Pasteur, 75015 Paris, France}
\affil[2]{Institute of Science and Technology Austria, Klosterneuburg, Austria}
\affil[3]{Joseph Henry Laboratories of Physics \& Lewis-Sigler Institute for Integrative Genomics,\protect\\ Princeton University, Princeton 08544, USA}
\affil[4]{D\'epartement de Biochimie et M\'edecine Mol\'eculaire, Universit\'e de Montr\'eal, Montr\'eal, QC H3C 3J7, Canada}
\affil[5]{Institut Courtois d'innovation Biom\'edicale, Montr\'eal, QC H2S 3H1, Canada}
\affil[6]{Mila -- Quebec AI Institute, Montr\'eal, QC H2S 3H1, Canada}

\begin{document}

\twocolumn[
\begin{@twocolumnfalse}

\maketitle

\begingroup
\renewcommand\thefootnote{}
\setlength{\footnotemargin}{0pt}
\footnotetext{$^{*}$Corresponding author: paul.francois@umontreal.ca}
\endgroup

\begin{center}
\begin{minipage}{0.85\linewidth}
\begin{abstract}
Understanding the space of ``the actual and the possible'' regulatory architectures capable of realizing biological functions is one of the major challenges in biology. Optimization-based approaches can rationalize regulatory mechanisms from functional objectives and constraints, but their computational cost has limited them to finding individual solutions. Here, we introduce \texttt{jaxgrn}, a framework that combines differentiable programming with stochastic simulations of gene regulatory networks (GRNs) to optimize thousands of networks end-to-end on GPUs. Using morphogen-driven embryonic patterning as a case study, we optimize networks for domain formation and positional information, while varying constraints on input noise, boundary precision and scaling with system size. While recovering known archetypes for scaling, we find that varying the constraints organizes optimal solutions into qualitatively distinct regimes. When optimizing domain formation, we observe phenotypic convergence across a continuum of regulatory architectures, ranging from input-driven to feedback-driven networks. When optimizing positional information, solutions are initially phenotypically diverse despite moderate variation in regulatory architecture; imposing scaling drives both phenotypic and regulatory convergence of the optimized networks onto a few recurring regulatory architectures. Our systematic approach opens the door for variational systems biology in which changing functional constraints reshape the landscape of possible phenotypes and regulatory architectures.
\end{abstract}
\end{minipage}
\end{center}

\vspace{0.6cm}

\end{@twocolumnfalse}
]

\section{Introduction}\label{introduction}

The origins of biological structures (from proteins to gene networks, organs and body-plan) remain one of the most fundamental questions of biology~\cite{wagner2015arrival, erwin2026origins}. While we know that Darwinian evolution is the engine of biological changes, there is much debate about how directional it is, and the question remains whether we can predict anything about biology. While some authors (such as Gould) have insisted on the contingent aspects of evolution~\cite{gould1989wonderful}, suggesting it is essentially unpredictable, there are also many examples at all scales of convergent evolution, suggesting that there are indeed regularities that can be rationalized and predicted~\cite{losos2011convergence}. Furthermore, the variations accessible to evolution are themselves highly structured: developmental and genotype-to-phenotype mappings can strongly bias which phenotypes are more readily produced than others~\cite{uller2018developmental,kaneko2007evolution}. Evolutionary dynamics can themselves restrict phenotypic variation to lower-dimensional manifolds~\cite{furusawa2018formation}. Convergence could therefore reflect not only selection for similar functions, but also constraints on the space of accessible solutions~\cite{schaerli2018synthetic}.

A classical (but debated) approach is to frame evolution as an optimization problem.
Focusing on cellular processing, if the environmental and internal constraints on the one hand, and the cellular function on the other hand can be formalized and identified, it should indeed be possible to infer biological mechanisms from first principles~\cite{bialek2012biophysics, tkacik_information_2008}. Such a perspective has already led to major insights into biology, for instance on the structural correspondence between biological networks and the ones independently designed by engineers~\cite{mangan2003structure}. Can we then characterize not only individual solutions to a given biological function, but the space of possible solutions and how it is shaped by different constraints? In other words, rather than a single optimum, we seek to characterize the \textit{ensemble} of solutions compatible with a given function and set of constraints~\cite{mlynarski_statistical_2021, bruckner_marr_2026}.

There are a few classical examples of evolutionary prediction through gradient descent, e.g.\ the derivation of a plausible evolutionary pathway from a photolayer of cells to a proper eye from purely physical considerations~\cite{nilsson1994pessimistic}. However, given the wide range of the ``actual and the possible''~\cite{jacob1977evolution}, computational methods are necessary to explore the optimization of complex phenotypes. Examples include directed network enumeration~\cite{cotterell2010atlas, schaerli2014unified}, evolutionary simulations~\cite{francois_design_2004, francois_predicting_2010, booth2025gene}, neural network optimization to match gene network behavior \cite{hiscock2019adapting,shen2021finding} or Monte Carlo/annealing simulations exploring how specific gene network structures naturally emerge to solve specific problems~\cite{sokolowski_deriving_2025}. These approaches have led to predictive and actionable insights into biology, e.g.\ in silico evolution of discrimination models for immunology suggested an Adaptive Kinetic Proofreading framework~\cite{lalanne2013principles}, validated experimentally and eventually leading to the development of new cancer immunotherapies~\cite{kondo2025engineering}. 

\begin{figure*}[tp]
\centering
\includegraphics[width=\textwidth,height=0.5\textheight,keepaspectratio]{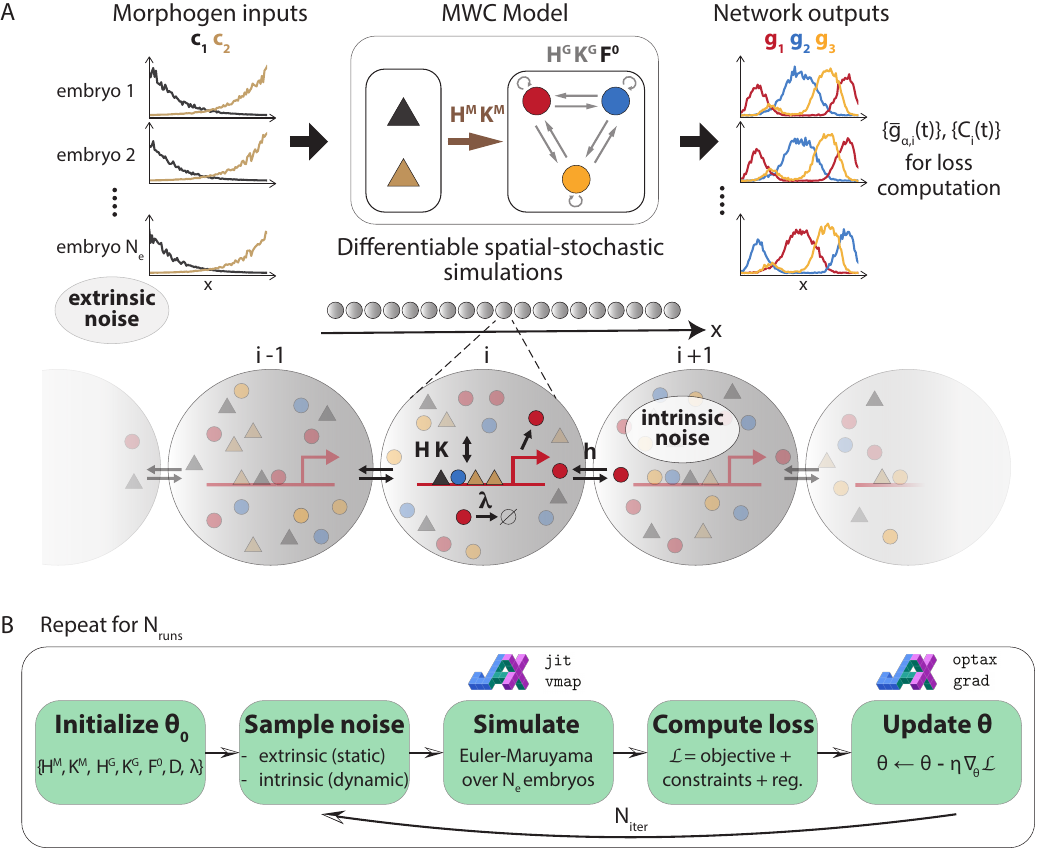}
\caption
    [\texttt{jaxgrn}: differentiable simulation and optimization of morphogen-driven GRN patterning.]
    {
    \textbf{\texttt{jaxgrn}: differentiable simulation and optimization of morphogen-driven GRN patterning.}
    {\bf A}:~Externally specified morphogen profiles drive a network of $N_g$ genes whose regulation is computed locally using an MWC model for gene regulation (Eq.~\ref{eq:mwc}) and converted into a production rate. 
    Gene products then evolve in 1D space (along the AP axis) and time according to discretized reaction-diffusion dynamics with decay and diffusive coupling between neighboring nuclei (Eq.~\ref{eq:pde_continuous}). Stochasticity is incorporated at two levels: embryo-to-embryo extrinsic variability is applied directly to morphogen inputs, while intrinsic fluctuations are introduced during time integration through a chemical Langevin approximation including production, degradation and diffusion hopping noise. Simulating an ensemble of embryos produces spatiotemporal expression patterns $\{g_{\alpha,i}^{(e)}(t)\}$ and their statistics (means and covariances), which are used to evaluate optimization objectives and biological constraints.
    {\bf B}:~At each training step, a parameter set $\theta$ (regulatory parameters and, optionally, physical parameters such as $D$ and $\lambda$, which are held fixed in this study) is used to generate an ensemble of stochastic gene expression trajectories by forward simulation. The resulting spatiotemporal profiles are mapped to a scalar loss consisting of a primary objective term together with biological constraints and regularization. 
    Gradients of the total loss with respect to $\theta$ are computed by automatic differentiation through the simulator and loss evaluation, enabling gradient-based parameter updates. Noise sources are resampled across iterations to favor robust solutions and independent initializations can be optimized in parallel to recover diverse ensembles of optimal solutions.
}
\label{fig:model_schematics_pipeline}
\end{figure*}

However, existing methods for gene network optimization remain limited in their ability to characterize the diversity of solutions to a given biological function. Previous approaches have identified individual solutions or explored relatively small sets of networks, leaving higher-level questions about the structure of solution spaces and their dependence on functional constraints largely unanswered. Recent advances in machine learning offer opportunities to overcome these limitations. GPU-based simulations combined with differentiable programming have enabled efficient optimization of artificial neural networks with billions of parameters~\cite{kaplan2020scaling}, and these approaches are increasingly being extended to biological models~\cite{alquraishi_differentiable_2021}, including detailed biophysical neuron models~\cite{deistler_jaxley_2025} and gene networks driving the morphogenesis of growing cell clusters~\cite{deshpande_engineering_2025}. 

Here, we introduce \texttt{jaxgrn}, a high-throughput framework that combines differentiable programming with stochastic simulations to optimize large ensembles of gene regulatory networks for specific biological functions. Using morphogen-driven embryonic patterning as a case study, we optimize networks for domain formation and positional information under varying constraints on input noise, boundary precision and scaling with system size. 
We show that constraints qualitatively reshape the distribution of optimized solutions, producing regimes ranging from phenotypic diversity to convergence in both phenotypic and regulatory properties. More generally, our results establish differentiable optimization as a practical approach for systematically investigating how functional constraints structure the space of possible regulatory architectures and phenotypes.

\section{Results}

\subsection{A JAX-based optimization pipeline for spatial stochastic patterning models of gene networks}

Our goal is to optimize families of gene networks for various properties through end-to-end automatic differentiation. Our implementation is organized as a modular pipeline separating (i) model definition, (ii) stochastic simulation, (iii) loss evaluation and (iv) gradient-based optimization. For a given gene-network parameter set $\theta$, we generate an ensemble of stochastic gene expression patterns. The simulated profiles are then mapped to scalar objectives and constraint penalties and the resulting scalar loss is minimized using gradient descent. The framework is modular at the level of parameterization, allowing different constrained or transformed representations of model parameters (e.g.\ bounded variables, masked couplings or sign-constrained interactions) without changing the simulator or loss definitions. 

Figure~\ref{fig:model_schematics_pipeline}A introduces both our biochemical model and the specific system we are considering. We generate gene expression patterns for an ensemble of $N_e$ simulated embryos on which to compute the optimization objectives. We simulate morphogen-driven reaction-diffusion dynamics in a spatial stochastic framework along a one-dimensional axis (biologically corresponding to the anterior-posterior (AP) axis). We consider $N_m$ morphogen inputs and $N_g$ network genes. We model gene expression via a network of genes, driven by external morphogen inputs, that evolves through regulation, decay, and diffusion between neighboring nuclei. Following~\cite{sokolowski_deriving_2025}, gene regulation is captured by a Monod-Wyman-Changeux (MWC) framework, in which each gene's activity is a sigmoidal function of a regulation potential combining weighted, saturable contributions from morphogens and other network genes, offering a more interpretable and flexible alternative to standard Hill functions. The optimized parameters $\theta$ are the strengths $H$ and dissociation constants $K$ of all morphogen--gene and gene--gene interactions, together with gene-specific activation thresholds $F^{(0)}$, i.e.\ 52 parameters for four genes and two morphogen inputs, and 60 for three inputs. We include both extrinsic variability (embryo-to-embryo fluctuations of the morphogen inputs, hereafter ``input noise'') and intrinsic variability due to the stochasticity of biochemical reactions, modeled by a chemical Langevin approximation of the underlying chemical master equation~\cite{gillespie_chemical_2000}. For any given network, we integrate the corresponding stochastic differential equations over an ensemble of embryos, using the explicit Euler-Maruyama method~\cite{kloeden_numerical_1992}, with a ReLU projection applied at each step to enforce non-negative expression levels (a small numerical correction for rare negative excursions due to noise) while using a straight-through estimator~\cite{bengio_estimating_2013} to preserve differentiability. More mathematical and implementation details are given in Methods (Sec.~\ref{sec:methods}) and in the Supplement. 

A key design choice is to treat the entire workflow as a single differentiable computation graph, enabling end-to-end optimization by automatic differentiation of regulatory parameters and, where desired, of dynamical parameters such as diffusion coefficients and decay rates (both held fixed in the case studies below, see \meth{sec:model}). We implement this framework in JAX~\cite{bradbury_jax_2018}, using just-in-time compilation and automatic vectorization to simulate large ensembles of stochastic embryos in parallel on GPU. This makes it practical to optimize stochastic dynamical systems directly from simulation-derived statistics and, in our setting, yields substantially faster convergence than gradient-free alternatives for the same model class and objective~\cite{sokolowski_deriving_2025}.

Parameter updates are performed using the Adam optimizer~\cite{kingma_adam_2017}, which minimizes a problem-specific loss function $\mathcal{L}_{\mathrm{total}}(\theta)$ (\meth{sec:loss} and Supplement) using gradients obtained by automatic differentiation through the simulation pipeline. At each optimization step, noise sources are resampled to prevent overfitting to a fixed stochastic realization and to encourage robust solutions. This implements an effective stochastic gradient descent in parameter space.

The pipeline is driven by a command-line interface that fully specifies the model class, noise settings and objective/constraint configuration, enabling reproducible sweeps across thousands of optimization runs. Each run produces a self-contained output folder containing the full configuration, model checkpoints, scalar logs and standardized diagnostic plots, facilitating systematic hyperparameter exploration and benchmarking across model classes and objective choices.

The full simulation-to-optimization pipeline and the main JAX transformations used are summarized in Fig.~\ref{fig:model_schematics_pipeline}B.

Below, we study the optimization of three phenotypes as case studies: two primary ones, and a secondary one (i.e.\ contingent on the primary ones). The two primary objectives optimized respectively are domain definition (similar to Hox gene patterning)~\cite{francois_predicting_2010} and positional information~\cite{sokolowski_deriving_2025}. More details are given in each section; both of those phenotypes were previously explored using Monte Carlo-based or evolutionary optimizations, but our framework allows for a much more extensive exploration of solutions. As a secondary objective, we are interested in the problem of scale-invariant patterning. Despite fluctuations in length and size of embryos, it has been observed both in vertebrates and invertebrates that gene expression patterns in development typically scale with size~\cite{cooke1981scale,gregor2005diffusion,ben2008scaling}, a property that also generalizes to stem cell systems and gastruloids~\cite{lauschke2013scaling,merle2024precise}. This has been most recently quantified in \textit{Drosophila} embryos, for the gap gene system: while the embryonic lengths $L$ fluctuate by $\pm 5\%$, the corresponding gap gene pattern scales almost perfectly~\cite{nikolic_scale_2024}, despite the fact that the morphogenetic inputs do not~\cite{gregor2005diffusion}. This suggests that scaling emerges through network interaction, and offers a good case study of the emergence of regulatory strategies implementing complex phenotypes.

\begin{figure*}[tp]
\centering
\includegraphics[width=\textwidth,height=0.68\textheight,keepaspectratio]{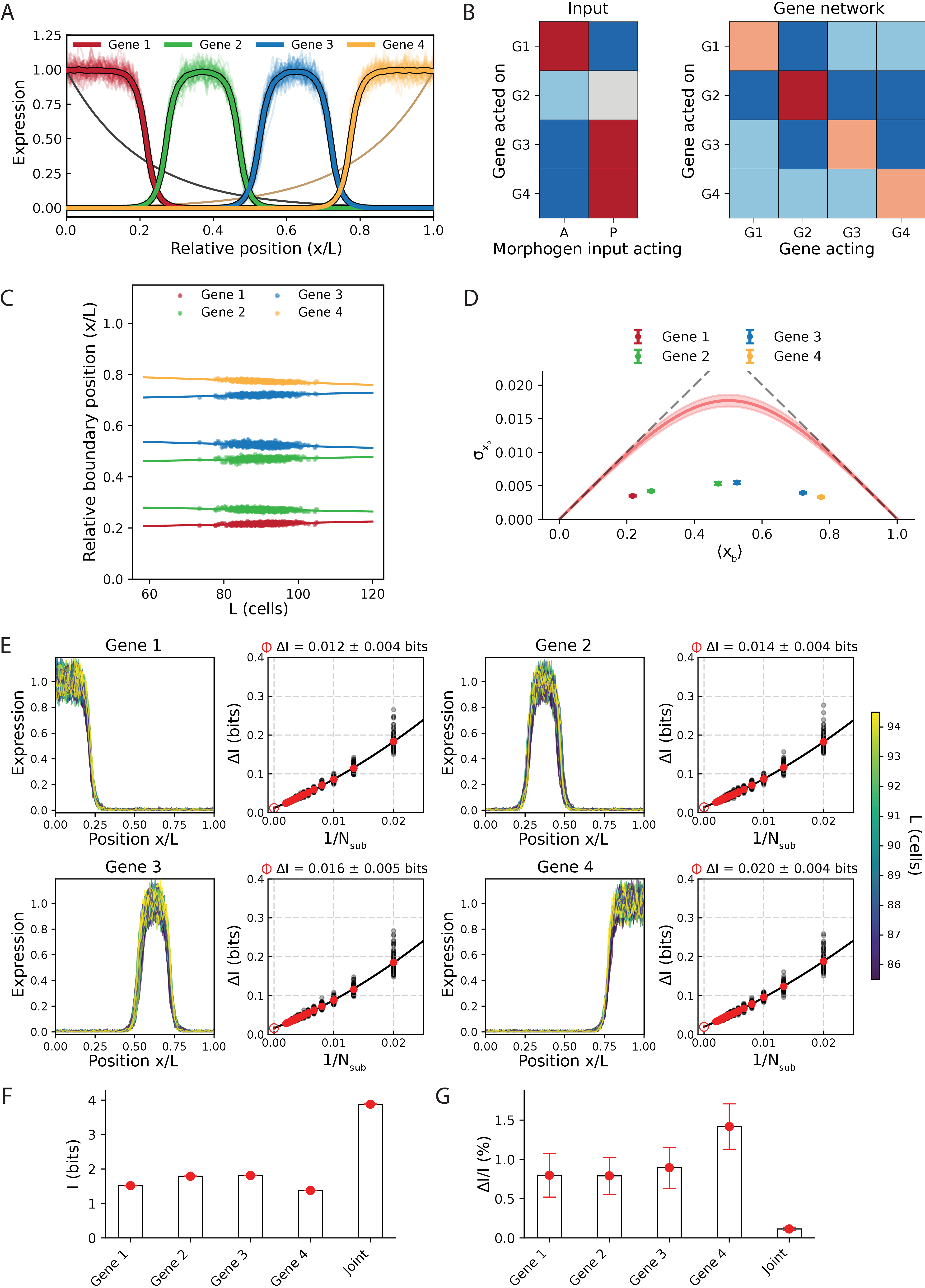}
\caption
    [Example of a fully scaling four-gene GRN.]
    {
    \textbf{Example of a fully scaling four-gene GRN.}
    Network optimized for domain formation (mutual-information objective) with the scaling constraint.
    {\bf A}:~Mean (bold) and single-embryo (light) expression profiles of the optimized network versus relative position; fixed morphogen inputs at reference length $L_0$ in background.
    {\bf B}:~Effective contributions \mbox{$W_{ij} = H_{ij} \log(1 + 1/K_{ij})$} of each regulator $i$ (morphogen or gene) on target gene $j$: negligible ($\lvert W_{ij}\rvert<1$, grey), weak ($1\leq\lvert W_{ij}\rvert<10$, light) or strong ($\lvert W_{ij}\rvert\geq10$, dark); red, activation; blue, repression. Unlike the other genes, G2 is barely regulated by the inputs but strongly repressed by all other genes.
    {\bf C}:~Relative boundary positions versus embryo length $L$, with linear fits.
    {\bf D}:~Standard deviation of boundary positions versus mean position (bootstrapped errors); red curve, expected precision for non-scaling boundaries anchored to the poles (Supplement).
    {\bf E}:~Single-embryo profiles colored by $L$ (left) and scaling deviation $\Delta I$ estimated from random subsets of $N_{\mathrm{sub}}$ embryos, extrapolated to $N_{\mathrm{sub}}\to\infty$ (right; Supplement).
    {\bf F, G}:~Positional information $I$ and fraction of information about absolute position $\Delta I/I$ for single genes and the joint pattern. Error bars in \textbf{G} are bootstrapped.
}
\label{fig:MI_scaling_example}
\end{figure*}

\subsection{Scaling in a minimal domain-formation problem} \label{domain_scaling}

The anterior-posterior axis of metazoans is organized into discrete regions, with sharp boundaries, specifying local identity (head, thorax, types of vertebrae...). Remarkably, those regions are defined by genes belonging to the Hox family, expressed in discrete, well-defined domains along the AP axis. The problem of gene network evolution optimizing such discrete organization has been previously studied using in silico evolution of gene networks, using a mutual-information metric that optimizes regional identities (high number of well-defined regions with specific genes expressed, see \meth{sec:domain_loss})~\cite{francois_predicting_2010}. As a first case study, we revisit this problem by optimizing domain formation across embryos with fluctuating lengths $L$.

We fixed the number of genes to $4$, so that optimizing mutual information between position and gene expression following~\cite{francois_predicting_2010} should give rise to 4 regions of equal size, where only one of the four genes is expressed. The challenge is to \textit{simultaneously} optimize this mutual information with scale-invariance, i.e.\ consistency of domain positions in relative coordinates across embryo sizes. As our objective we use the mutual information in relative coordinates between position and cell-fate identity to reward domain formation of equal size~\cite{francois_predicting_2010} (\meth{sec:domain_loss}). Because this objective is evaluated in relative coordinates and pooled over embryos of different lengths, it implicitly promotes scaling: boundaries that shift in relative coordinates with $L$ blur fate assignments across the ensemble and lower the mutual information.

We optimized GRN models under the control of two opposing non-scaling morphogen inputs: an anterior gradient with absolute decay length $\ell_A$ and a posterior gradient with absolute decay length $\ell_P$, as defined in Eq.~\ref{eq:non_scaling_gradient}, so that neither gradient scales with length. In this minimal domain-formation problem, morphogen decay lengths are fixed (not optimized), and we set the input noise to zero ($\sigma_{\mathrm{IN}}=0$), retaining intrinsic expression noise and embryo-length variability. Optimizing the mutual-information objective alone, however, rarely yields fully scaling four-gene solutions (Supplementary Figures). We therefore add an explicit scaling constraint based on the mean absolute deviation (MAD), which measures the average deviation of individual-embryo expression profiles from the ensemble mean profile in relative coordinates, and is thus small when profiles of embryos of different lengths superimpose (\meth{sec:scaling}).

Fig.~\ref{fig:MI_scaling_example} shows an illustrative example of a fully scaling GRN, optimized through our framework, displaying high precision and scale-invariant boundaries. Panel A shows the profiles of both input morphogens and downstream  4 genes after optimization, illustrating the four well-separated domains along the relative position. Fig.~\ref{fig:MI_scaling_example}B displays the corresponding interaction matrix strength between genes (see details on its computation in \meth{sec:network}; red corresponds to activation and blue to repression; darker colors correspond to stronger contributions at maximal expression). Of note, only three of the four genes are significantly regulated by both morphogens, and we observe an interaction pattern in which all genes self-activate and mutually repress.

Fig.~\ref{fig:MI_scaling_example}C--G provide quantitative measures of scaling properties. Fig.~\ref{fig:MI_scaling_example}C shows the relative positions of the domain boundaries as a function of embryo length. Lines are nearly flat: residual slopes are small but nonzero, corresponding to boundary shifts below one cell over the range of simulated lengths, within our scaling criteria (\meth{sec:scaling}). More quantitatively, Fig.~\ref{fig:MI_scaling_example}D shows the standard deviation of boundary positions shown in \textbf{C}, as a function of the mean boundary position, with typical values around $0.5\%$ of the embryonic length (to be compared to the $5\%$ variation in embryonic length we impose). For further comparison, we also show an estimate of precision (red line in Fig.~\ref{fig:MI_scaling_example}D) based on the limiting case of non-scaling anchored boundaries (this estimate depends on relative position and vanishes at $x=0$ and $x=1$, since the poles are well defined irrespective of embryo length). So the optimized standard deviation is well below this natural limit, as expected for scaling solutions.

Finally, we use an information-based measurement, the deviation from scale invariance $\Delta I$, which quantifies how much positional information is gained by knowing the length of the embryo (see \meth{sec:scaling} for definition, $\Delta I=0$ corresponds to perfect scaling). Fig.~\ref{fig:MI_scaling_example}E shows gene expression profiles as a function of relative position for each gene and a range of embryo lengths, as well as estimates of $\Delta I$. Visually, we see that rescaled profiles collapse on one another irrespective of the length, indicating scaling, which is quantitatively confirmed by the very low $\Delta I$ for all four genes. The joint pattern carries about 4 bits of positional information (Fig.~\ref{fig:MI_scaling_example}F), and Fig.~\ref{fig:MI_scaling_example}G quantifies the fraction of information about absolute position for each gene and shows that it is smaller than $2\%$ both for individual genes and for the joint distribution (thus satisfying the scaling criteria defined in \meth{sec:scaling}). Thus, this example illustrates that our optimization method allows for the optimization of scaling domain solutions using information-based criteria, with clear, quantitative measurements of information content, position and scaling.

\begin{figure*}[tp]
\centering
\includegraphics[width=\textwidth,height=0.8\textheight,keepaspectratio]{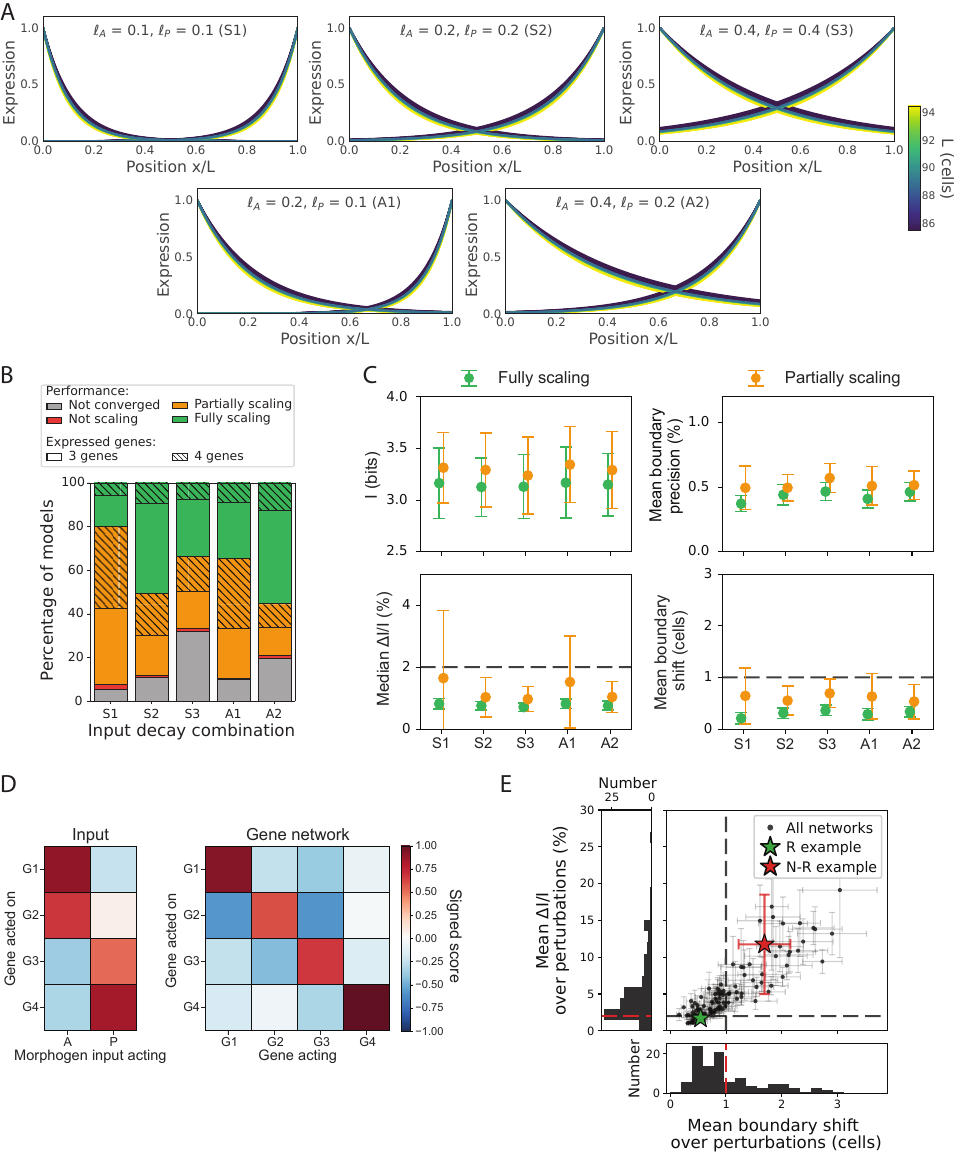}
\caption
    [Optimization of scaling GRNs under different non-scaling input scenarios in a minimal domain-formation problem.]
    {
    \textbf{Optimization of scaling GRNs under different non-scaling input scenarios in a minimal domain-formation problem.}
    Panels B--D show results obtained with the scaling constraint; corresponding results without it are shown in Supplementary Figures.
    {\bf A}:~Symmetric (S, top) and asymmetric (A, bottom) opposing input scenarios, with decay lengths $\ell_A$ and $\ell_P$ in units of $L_0$.
    {\bf B}:~Optimization outcomes by scaling performance and number of expressed genes.
    {\bf C}:~Positional information $I$, mean boundary precision, $\Delta I/I$ (median over genes) and mean boundary shift $\Delta x_b$ for fully and partially scaling GRNs. Mean $\pm$ standard deviation across networks; dashed lines, scaling thresholds.
    {\bf D}:~Signed-score interaction matrix of the 88 fully scaling four-gene GRNs, genes ordered anterior to posterior.
    {\bf E}:~Perturbation-averaged $\Delta I/I$ versus $\Delta x_b$ for the 107 fully scaling four-gene GRNs (19 optimized without and 88 with the scaling constraint), with marginal histograms. Error bars, standard error across perturbations; dashed lines, scaling thresholds. Green and red stars, robust (Fig.~\ref{fig:MI_scaling_example}) and non-robust (Supplementary Figures) examples.
}
\label{fig:MI_figure1_general_results}
\end{figure*}

\subsection{A statistical analysis of domain-scaling solutions}

Having established that our framework can find working solutions, we now test if we can generate extensive families of solutions, with multiple scenarios and enough statistical power to disentangle the effect of multiple constraints on the optimization of domain scaling.

\subsubsection{Morphogen shapes}
We first tested the influence of the shapes of the morphogen inputs. We considered various symmetric and asymmetric scenarios with different combinations of $\ell_A$ and $\ell_P$ to vary gradient steepness and the range of possible positional readouts (Fig.~\ref{fig:MI_figure1_general_results}A), and optimized 200 GRNs per input combination with the scaling constraint. We then categorized the optimization outcomes: GRNs in which no pattern formed, or for which optimization did not converge, are distinguished from models that scale either fully or partially. Partially scaling GRNs have at least one gene but not all satisfying the scaling criteria (\meth{sec:scaling}).
 
Depending on the input scenario, between 20 and 55\% of networks fully scale, with either three or four expressed genes, with intermediate-steepness scenarios A2 and S2 yielding the highest fractions of fully scaling models (Fig.~\ref{fig:MI_figure1_general_results}B). This result is intuitive: we expect scaling to be most difficult far from the poles, where morphogen levels are low, and S2 corresponds to the case where both morphogens are still present in the center, with substantial relative variation with respect to one another. Positional information, boundary precision and scaling metrics ($\Delta I/I$ and boundary shift $\Delta x_b$, \meth{sec:scaling}) are summarized in Fig.~\ref{fig:MI_figure1_general_results}C. There is a clear difference between partially scaling and scaling solutions as expected but, interestingly, for the scaling solutions and boundaries, we do not see much difference between input scenarios.

For comparison, we also performed 200 optimizations per input scenario with the mutual-information objective alone (Supplementary Figures). Adding the scaling constraint increased the overall fraction of fully scaling solutions from 31\% to 39\%, and the fraction of fully scaling four-gene solutions from 2\% to 9\%.

\subsubsection{Phenotypic and regulatory convergence}

Among fully scaling solutions, we observe limited phenotypic variability, i.e.\ all fully optimized patterns present at least 3 domains of equal size (as expected from the loss function used~\cite{francois_predicting_2010}). Such convergence in phenotypes (i.e.\ pattern) raises the question of the underlying genotypes (i.e.\ parameters) and corresponding regulatory strategies (i.e.\ correlated interaction signs and strengths). We computed a signed-score interaction matrix from the 88 fully scaling four-gene solutions obtained by optimizing mutual information with an explicit scaling constraint, pooling all five input scenarios and ordering the genes from anterior to posterior (Supplement); this matrix is shown in Fig.~\ref{fig:MI_figure1_general_results}D. The pattern we find is remarkably simple and intuitive. There is first a graded regulation pattern by the inputs (anterior to the left), with the leftmost gene (G1) strongly activated by the anterior morphogen (input 1) and weakly repressed by the posterior one (input 2). The second gene from the left (G2) is moderately activated by input 1 and only marginally regulated by input 2. We observe a symmetrical situation for the two rightmost (posterior) genes. All genes further self-activate, more strongly for genes at the boundaries, and neighboring genes repress one another. Mutual repression between neighboring genes is known to sharpen boundaries and improve their precision~\cite{sokolowski_mutual_2012}, and interactions among target genes can increase the information they jointly transmit about their inputs~\cite{walczak_optimizing_2010}; but our high-throughput approach allows for the establishment of more general rules, with subtle effects on the magnitude of interactions as a function of position. Notice also that, while average interactions are well defined, the example shown in Fig.~\ref{fig:MI_scaling_example} presents both typical features (neighboring genes strongly repressing) and exceptions to these rules (regulation of Gene $2$), showing that, even for relatively simple problems, there is strong variability in the solutions found.

\subsubsection{Exploring regulatory strategies through perturbation analysis}

Having established general features of scaling solutions, we can also use our framework to study which aspects of the networks found matter the most for the scaling behavior. Inspired by biology, we followed a targeted mutation approach with two classes: (i) single gene knockouts (KO) for each of the four genes and (ii) regulatory interaction perturbations, selectively removing contributions from the matrix $H^G$ by setting selected classes of interaction strengths to zero. As in a laboratory knockout, all other parameters are kept at their optimized values, without re-optimization~\cite{sokolowski_deriving_2025}. Specifically, we removed auto-regulation, nearest-neighbor (NN) interactions or all network interactions, and also tested networks retaining only auto-regulation, only NN interactions, or both, the latter removing all next-nearest-neighbor and longer-range interactions (\meth{sec:network}). These perturbations assess the relative roles of morphogen inputs versus network interactions in establishing scale-invariant boundaries.

We first examined the example GRN from Fig.~\ref{fig:MI_scaling_example} (details in Supplementary Figures). Scaling in this network is robust to both removal of genes and removal of interactions, with relative domain positions largely preserved. This suggests that scaling is input-driven, particularly through the regulation of genes 1, 3 and 4. In the Supplement, we show how this can occur: thresholds defined by linear combinations of the logarithms of the opposing morphogen concentrations can specify boundaries at fixed relative positions as embryo length varies, consistent with earlier work on combinatorial decoding of opposing gradients~\cite{mchale_embryonic_2006,shen_scaling_2022}.

We contrasted this with another GRN from scenario A1 with weaker morphogen contributions but preserved scale invariance (Supplement). There, the central genes are significantly less robust to perturbations, as their boundary scaling depends more on network interactions than morphogen inputs. In particular, perturbations that remove nearest-neighbor regulatory interactions are the most detrimental to scaling. This is reminiscent of previously proposed architectures based on mutual repression between neighboring domains~\cite{sokolowski_mutual_2012}.

To test whether these behaviors are general features of scaling solutions to the domain formation problem, we repeated the perturbation analysis across all 107 fully scaling four-gene GRNs, pooled across the five input scenarios and across optimizations with and without the scaling constraint (Fig.~\ref{fig:MI_figure1_general_results}E). For each GRN and perturbation type, we computed the fraction of information about absolute position $\Delta I/I$ (near zero for scale invariance) and the mean boundary shift $\Delta x_b$, and then averaged over perturbations. We see a continuum of solutions, with strong correlations between the two metrics: when a perturbation breaks scaling, it does so by making boundaries shift systematically with embryo length, which in turn makes the pattern informative about absolute position. Robust GRNs, such as the input-driven example of Fig.~\ref{fig:MI_scaling_example} (green star), still satisfy both scaling criteria on average after perturbation, whereas non-robust GRNs, such as the network-driven example from scenario A1 (red star), violate both. This establishes that, even if we found multiple scaling solutions with almost exactly the same phenotype, there is degeneracy not only in the genotype space (strength of interactions), but also in the higher-level regulatory space (overall strategy underlying scaling).

\begin{figure*}[tp]
\centering
\includegraphics[width=\textwidth,height=0.7\textheight,keepaspectratio]{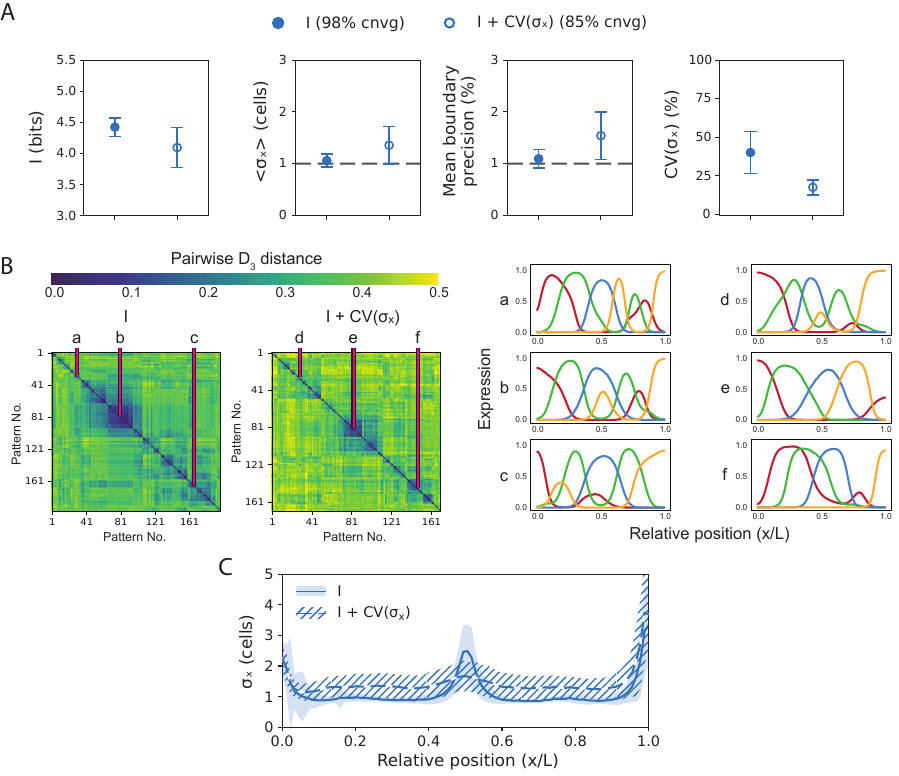}
\caption
    [Without embryo-length variability, a constraint promoting uniform positional error reshapes precision while preserving pattern diversity.]
    {
    \textbf{Without embryo-length variability, a constraint promoting uniform positional error reshapes precision while preserving pattern diversity.}
    Networks optimized and evaluated without embryo-length variability (all embryos of length $L_0$), with $\sigma_{\mathrm{IN}}=0.15$; 200 initializations per scenario.
    {\bf A}:~Positional information $I$, mean positional error $\langle\sigma_x\rangle$, mean boundary precision and $\mathrm{CV}(\sigma_x)$ without ($I$) and with ($I+\mathrm{CV}(\sigma_x)$) the uniformity constraint. Mean $\pm$ standard deviation across converged networks ($I\ge3$ bits; convergence percentages in legend).
    {\bf B}:~Pairwise $D_3$ distance matrices of converged networks (left), clustered independently for each scenario. Vertical lines in the distance matrices indicate representative patterns shown on the right. These examples illustrate the qualitative diversity of spatial expression profiles obtained across independent optimization runs.
    {\bf C}:~Positional error $\sigma_x$ along the AP axis: mean across converged networks (solid, $I$; dashed, $I+\mathrm{CV}(\sigma_x)$) $\pm$ standard deviation (shading, hatching).
}
\label{fig:PI_LV0}
\end{figure*}

\subsection{Optimizing positional information}

We now move to a problem with a more complex function where the space of phenotypes should be much more degenerate. Inspired by studies of the \textit{Drosophila} gap gene system~\cite{petkova_optimal_2019,dubuis_positional_2013}, we generate gene networks optimizing mutual information between position and overall gene expression $I(\mathbf g; x)$, i.e.\ positional information (\meth{sec:pi}). This contrasts with the previous section in the sense that we no longer require a single gene to be expressed at any given position: rather, we expect that combinations of genes encode local information, allowing for a cell to decode its position from such combinations. Again, we consider optimization of gene networks with four genes (by analogy with the four main gap genes \textit{hb}, \textit{Kr}, \textit{kni} and \textit{gt}), and consider optimization in response to three input genes: one anterior (similar to \textit{bicoid}), one posterior (similar to \textit{caudal}), and one expressed at both ends (similar to the terminal \textit{torso} system); following \textit{Drosophila} nomenclature, these are labeled as maternal inputs A, P and T in the figures. To match the \textit{Drosophila} case, we further introduced extrinsic fluctuations in the inputs (\meth{sec:model}).

Of note, this optimization requires special care in the definition of the loss function. The standard mutual information estimator involves a Gaussian-mixture marginal entropy that lacks a closed-form expression~\cite{huber_entropy_2008}. To obtain a closed-form, differentiable approximation of positional information suitable for gradient-based optimization, we adapt an existing second-order entropy expansion for Gaussian mixtures~\cite{huber_entropy_2008} to serve as a training objective within our end-to-end differentiable pipeline. In detail, we rely on multivariate Gaussian approximations to fit the average gene profiles, their covariance, and Taylor-expand the marginal entropy
$S[P(\mathbf g)]$~\cite{huber_entropy_2008} that depends only on the average of gene profiles and the covariances (\meth{sec:pi}). Those can be fitted through differentiable functions, thus providing us with estimates of mutual information that are fully differentiable.

Using this differentiable positional information approximation $I(\mathbf g; x)$, we optimize GRN models via gradient descent. Following~\cite{sokolowski_deriving_2025}, we impose biologically informed sign constraints on regulatory couplings and ``resource constraints'' mimicking what is observed in \textit{Drosophila} (Methods, Secs.~\ref{sec:model} and~\ref{sec:loss}).

We tested optimization scenarios with different biologically relevant constraints, optimizing 200 models for each scenario from random initializations. We found that our optimization pipeline reliably discovered high positional information solutions (about 4.4 bits on average, Fig.~\ref{fig:PI_LV0}A, comparable to the 4.2 bits found in the \textit{Drosophila} gap gene system \cite{dubuis_positional_2013,sokolowski_deriving_2025}), with many different patterns (a sample is shown in Fig.~\ref{fig:PI_LV0}B). To quantify the diversity of solutions, we computed pairwise permutation-invariant $D_3$ distances between patterns (\meth{sec:network}). Pairwise $D_3$ distance matrices (Fig.~\ref{fig:PI_LV0}B) reveal structured regions of low pairwise distance, forming large clusters along the diagonal that frequently correspond to symmetric patterns (for example profile e in Fig.~\ref{fig:PI_LV0}B).

\subsubsection{Changing constraints modulates the nature and diversity of accessible solutions}

From this diverse baseline of network solutions optimizing high information content, we can study how various constraints can be added to shape the space of the solutions. 

Adding a precision-related constraint such as uniformity of the positional error $\sigma_x$ (the precision with which position can be decoded from gene expression, \meth{sec:pi}), quantified by its coefficient of variation $\mathrm{CV}(\sigma_x)$ along the AP axis, reshapes the high-information solution manifold. With such a constraint, we consistently obtained GRN models with more uniform positional error while maintaining comparably high positional information (Fig.~\ref{fig:PI_LV0}A,C). We note a trade-off, with improved uniformity accompanied by a modest reduction in mean positional information and an increase in mean positional error. Substantial heterogeneity in pattern space persisted across independently optimized high-information solutions, and the distributions of pairwise $D_3$ distances overlap across optimization scenarios, with comparable means and spread (Supplementary Figures). This suggests that the positional-error uniformity constraint restricts the set of high-information solutions but does not collapse it in pattern space.

We also varied the noise level in the inputs (Supplementary Figures). We found that pairwise $D_3$ distances decreased with increasing input noise during optimization, consistent with an implicit regularization effect that compressed the accessible high-performing manifold in pattern space, but without changing it qualitatively. At the GRN level, increasing optimization noise systematically altered the regulatory architecture, from more feedforward to more feedback-based architectures. These interaction changes paralleled the improved robustness to input noise, consistent with internal gene--gene coupling reducing effective sensitivity to fluctuations in maternal inputs. 

From these various case studies, we conclude that the differentiable approximation of positional information $I(\mathbf g;x)$ provides a stable objective for gradient-based optimization through spatial stochastic simulations. Our high-throughput optimizations allow us to show how additional constraints can modulate the space of possible solutions found, with more feedback-based regulation when input noise is increased.

\begin{figure*}[tp]
\centering
\includegraphics[width=\textwidth,height=0.7\textheight,keepaspectratio]{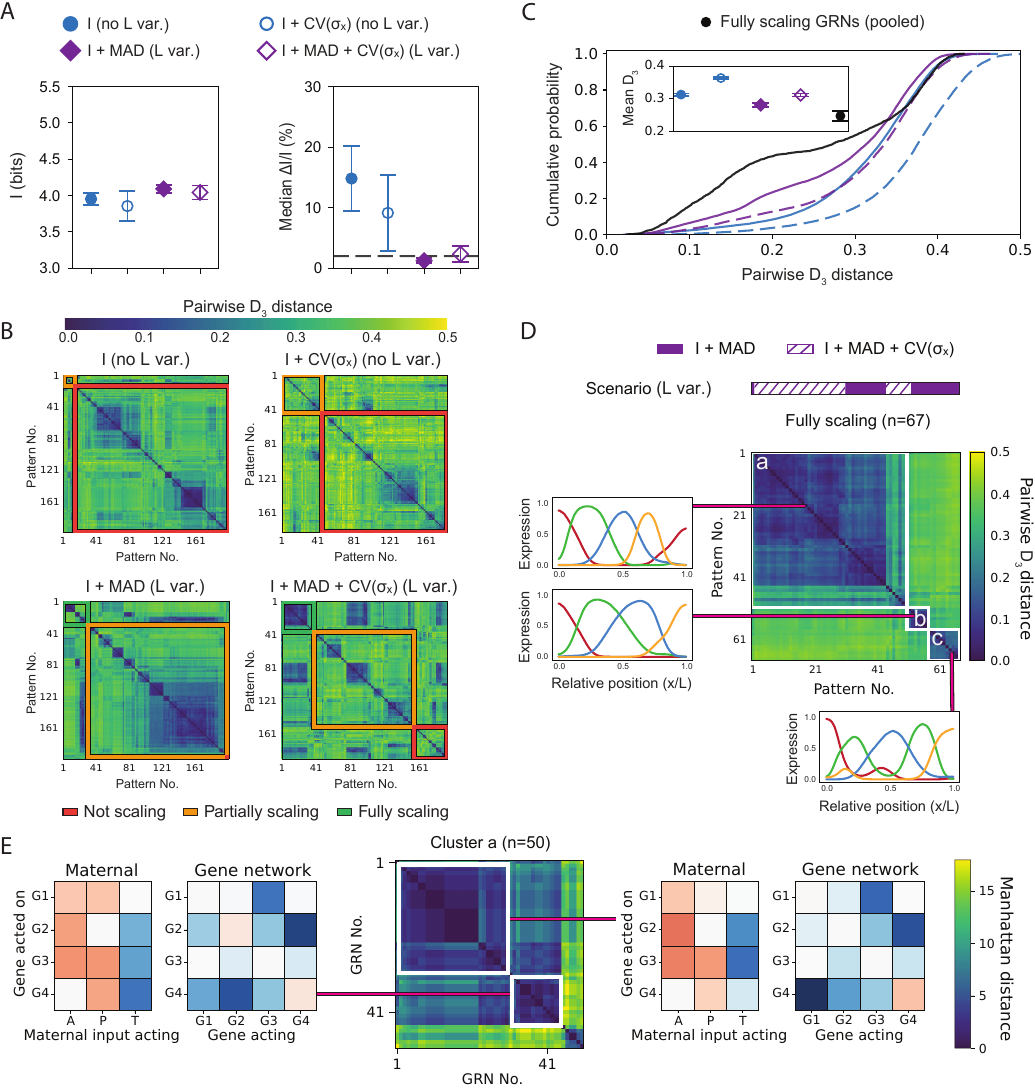}
\caption
    [With embryo-length variability, pattern diversity is reduced and regulatory structure converges among fully scaling high-positional-information GRNs.]
    {
    \textbf{With embryo-length variability, pattern diversity is reduced and regulatory structure converges among fully scaling high-positional-information GRNs.}
    Networks optimized at $\sigma_{\mathrm{IN}}^{\mathrm{opt}}=0.1$ and evaluated at $\sigma_{\mathrm{IN}}^{\mathrm{eval}}=0.15$ with $5\%$ length variability; 200 initializations per scenario.
    {\bf A}:~Positional information $I$ and median (over genes) $\Delta I/I$ for four scenarios: $I$ and $I+\mathrm{CV}(\sigma_x)$ optimized without length variability (``no $L$ var.''), $I+\mathrm{MAD}$ and $I+\mathrm{MAD}+\mathrm{CV}(\sigma_x)$ optimized with length variability (``$L$ var.''); all evaluated with length variability. Mean $\pm$ standard deviation across networks.
    {\bf B}:~Pairwise $D_3$ distance matrices per scenario, grouped by scaling performance: fully (green), partially (orange) or not scaling (red).
    {\bf C}:~Cumulative distributions of pairwise $D_3$ distances for the scenarios in \textbf{B} (colors as in \textbf{A}; dashed, with uniformity constraint) and for pooled fully scaling networks (black). Inset: mean $D_3$, with bootstrap standard errors from resampling networks within scenarios.
    {\bf D}:~$D_3$ matrix of fully scaling GRNs pooled across MAD scenarios ($n=67$; top band, scenario). White boxes, main pattern clusters a--c, with representative profiles (cluster medoids).
    {\bf E}:~Center: pairwise Manhattan distances between regulatory topology vectors within cluster~a ($n=50$; \meth{sec:network}); white boxes, two largest subclusters. Left and right: their mean regulatory matrices (red, activation; blue, repression). Genes ordered anterior to posterior.
}
\label{fig:PIScaling_diversity_architecture}
\end{figure*}

\begin{figure*}[tp]
\centering
\includegraphics[width=\textwidth,height=0.72\textheight,keepaspectratio]{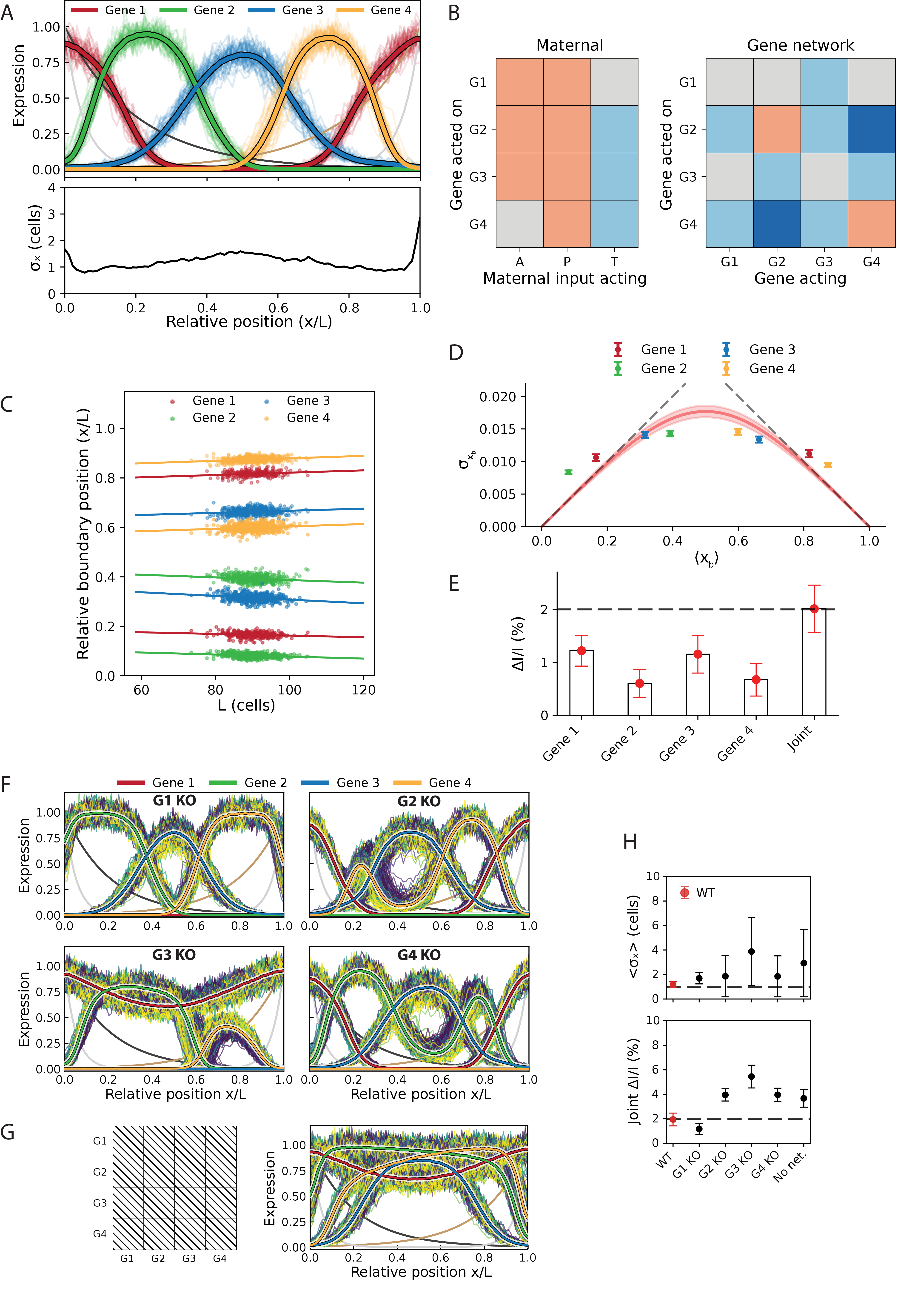}
\caption
    [Example of a fully scaling high-positional-information GRN.]
    {
    \textbf{Example of a fully scaling high-positional-information GRN.}
    Network optimized for positional information with scaling and positional-error uniformity constraints ($I+\mathrm{MAD}+\mathrm{CV}(\sigma_x)$).
    {\bf A--D}:~As Fig.~\ref{fig:MI_scaling_example}A--D; \textbf{A} additionally shows the positional error $\sigma_x$ along the AP axis. {\bf E}:~As Fig.~\ref{fig:MI_scaling_example}G.
    {\bf F}:~Single-gene knockouts (KO): single-embryo profiles for embryos of different lengths, with mean expression.
    {\bf G}:~Removal of all gene--gene interactions (No net., $H^G=0$, hatched), profiles as in \textbf{F}; other network perturbations in Supplementary Figures.
    {\bf H}:~Mean positional error and joint $\Delta I/I$ for wild type (WT) and the perturbations in \textbf{F} and \textbf{G}. Dashed lines, the reference positional error of one cell and the scaling threshold of $2\%$; error bars, standard deviations.
}
\label{fig:PIScaling_example}
\end{figure*}

\subsubsection{Pattern diversity is reduced among fully scaling high-positional-information networks}

The persistence of pattern diversity under the positional-error uniformity constraint and input-noise variations raises the question of whether such degeneracy survives a qualitatively different systems-level constraint: scale invariance. We combined the positional-information objective with the MAD scaling constraint introduced above ($I+\mathrm{MAD}$), with or without an additional constraint on the uniformity of positional error (Fig.~\ref{fig:PIScaling_diversity_architecture}A; see Methods, Secs.~\ref{sec:pi} and~\ref{sec:scaling}). Note that in contrast with the domain-formation problem, simply optimizing positional information over an ensemble of embryos of variable length is not sufficient to obtain scaling solutions: the pressure towards scale invariance carried by the objective is weaker and must be reinforced explicitly by the scaling constraint (Supplementary Figures). Since training at lower input noise yielded more fully scaling solutions (Supplementary Figures), networks in this section were optimized at $\sigma_{\mathrm{IN}}^{\mathrm{opt}}=0.1$ and evaluated at $\sigma_{\mathrm{IN}}^{\mathrm{eval}}=0.15$. For each scenario, we found that about 20\% of solutions were fully scaling. We also checked that networks optimized retain high positional information and comparable positional precision, while substantially reducing the scaling deviation $\langle\Delta I\rangle$, the fraction of information about absolute position $\langle\Delta I/I\rangle$ and the mean boundary shift $\langle\Delta x_b\rangle$ (Supplementary Figures). This shows that scale invariance can be selected directly without losing patterning performance: all four scenarios reach about 4 bits of positional information, while only MAD-optimized networks bring the median $\Delta I/I$ below threshold (Fig.~\ref{fig:PIScaling_diversity_architecture}A).

However, in contrast to the positional-error uniformity constraint and input-noise variations considered above, the scaling constraint strongly restricts the accessible pattern space, selecting a narrow subset within the broader high-information manifold. In particular, $D_3$ matrices show stronger clustering and reduced pairwise distances, with cumulative distributions systematically shifted toward much smaller values (see Supplementary Figures for comparisons of scaling metrics under all scenarios). When pooled across scenarios, the fully scaling GRNs (green sub-squares in Fig.~\ref{fig:PIScaling_diversity_architecture}B) occupy a more restricted region of pattern space than the full set of optimized solutions, with the lowest mean pairwise $D_3$ distance of all ensembles (Fig.~\ref{fig:PIScaling_diversity_architecture}C, black, and inset). Their pairwise distances reveal only a few correlated clusters, most of which correspond to approximately symmetric expression patterns. The very dominant cluster a (50 networks, Fig.~\ref{fig:PIScaling_diversity_architecture}D) bears resemblance to the sequential \textit{Kr}, \textit{kni} and posterior \textit{gt} domains sandwiched between two \textit{hb} regions, observed along the \textit{Drosophila} AP axis, with moderate overlaps between the four genes. We note that these patterns were already present in the positional-information optimization results without the scaling constraint (Supplementary Figures).

\subsubsection{Regulatory convergence for scaling with high-positional information}

To examine whether the \textit{phenotypic} convergence in pattern space reflects a corresponding \textit{regulatory} convergence in network space, we characterized the interactions of fully scaling GRNs within each pattern cluster (Fig.~\ref{fig:PIScaling_diversity_architecture}E). Pairwise Manhattan distances between topology vectors (\meth{sec:network}) reveal only a small number of tight subclusters, indicating that only a few distinct regulatory strategies are obtained. The dominant subcluster within each pattern group accounts for the majority of solutions and, in all cases, is characterized by weak repressive interactions between neighboring genes and strong repressive interactions between non-adjacent genes along the AP axis. Scaling thus acts as a strong filter on interactions, selecting a specific class of underlying regulatory mechanisms. This architecture corresponds to the ``alternating cushions'' scheme previously proposed for the \textit{Drosophila} gap gene system~\cite{vakulenko_size_2009, jaeger_dynamical_2004}.

We show one representative fully scaling GRN in Fig.~\ref{fig:PIScaling_example} from cluster a identified in Fig.~\ref{fig:PIScaling_diversity_architecture}D, together with its associated precision and scaling metrics. To get a better sense of how scaling works, we further performed perturbation analyses on representative scaling networks obtained (Fig.~\ref{fig:PIScaling_example} and see another example in the Supplementary Figures). Perturbation analysis showed that these repressions between non-adjacent genes are important for maintaining both precision and scale invariance: for instance, knocking out G2 (resp.\ G4) causes G4 (resp.\ G2) to expand into the vacated domain; this expression is still relatively precise (Fig.~\ref{fig:PIScaling_example}H) but no longer scales (Fig.~\ref{fig:PIScaling_example}F). The more central the gene, the larger the impact: for instance removal of the central G3 strongly perturbs both precision and scaling for all genes, while removal of G1 is much less impactful on both properties (Fig.~\ref{fig:PIScaling_example}F--H). In the Supplement, we further tested other perturbations, e.g.\ showing that while a minimal level of diffusion is necessary, diffusion plays no significant role in maintaining either precision or scale invariance in these networks.

\section{Discussion}

We propose a forward, predictive approach by leveraging modern machine learning tools to perform systematic optimizations of models of gene networks, under various constraints. We focused on embryonic patterning under the control of static morphogens as a case study, and optimized complex properties by gradient descent, such as scaling or positional information. Over multiple rounds, our approach produced thousands of solutions, with different properties depending on the constraints, which allows us to perform a detailed analysis of both the solutions themselves and their statistics.

At a granular level, we optimized networks for scaling, under multiple primary objectives, and, remarkably, recovered regulatory strategies related to three previously proposed archetypes: input-based decoding~\cite{mchale_embryonic_2006,shen_scaling_2022}, mutual repression between neighboring domains~\cite{sokolowski_mutual_2012}, and the second-neighbor-based ``alternating cushions'' architecture, with weak repression between adjacent domains and strong repression between next-adjacent ones~\cite{vakulenko_size_2009, jaeger_dynamical_2004}. While this validates the tool for enumerating plausible mechanisms, the diversity of solutions found rather suggests a continuous space of solutions, where the optimized networks tinker and combine several archetypes to get a scaling solution. Here we imposed the $\pm5\%$ length variability measured in \textit{Drosophila}; for such small variations, scaling only needs to hold to linear order in $\delta L/L$, which many architectures can achieve. We expect larger size variations to require scaling beyond linear order and thus to restrict solutions to even fewer strategies, a prediction directly testable within our framework.

Thus, beyond the specific cases studied, our high-throughput approach allows us to systematically probe the space of solutions for complex biological problems, shedding new light on some long-debated questions on the ``actual and the possible''~\cite{jacob1977evolution}. We first observe a rather complex genotype-to-phenotype mapping, best illustrated with the domain scaling problem, where the exact same pattern of four equally sized scaling domains is realized through a continuum of underlying network architectures in network space. Depending on the problem, we also get a spectrum of properties on the family of solutions themselves. When optimizing for positional information only, we get a zoo of solutions, with very comparable positional information, thus suggesting a highly diverse space of solutions for that problem specifically, consistent with previous work \cite{sokolowski_deriving_2025}. Conversely, when optimizing for scaling and information simultaneously, we see a collapse onto a few solutions. Interestingly, the dominant pattern found in the latter case was also observed when optimizing for positional information only, and gradually becomes more frequent as we increase the number of constraints such as precision. It thus seems that adding constraints does not simply improve or degrade performance, rather they reshape the space of accessible solutions. 

This naturally raises the question of why some specific solutions are obtained, especially in the latter regime where only very few emerge from gradient descent.
The simplest explanation is that the space of optimal solutions itself shrinks as more constraints are added, so that our optimization simply reflects this restriction. Alternatively, as the problem becomes more constrained (e.g.\ from information only to information + scaling), directions in network space might themselves become restricted, as previously observed with evolutionary approaches on simpler problems~\cite{franccois2007deriving, booth2025gene}, and more generally proposed from the emergence of low-dimensional or ``soft'' modes in evolved biological systems~\cite{furusawa2018formation, husain2020physical}. Only a few solutions might then be easily reachable by gradient descent, while others remain isolated or difficult to access, somewhat reminiscent of the ``survival of the flattest'' mechanism~\cite{wilke2001evolution}.

We did not disentangle the biases of gradient descent from the intrinsic structure of the solution space. Gradient-free Monte Carlo \cite{sokolowski_deriving_2025} and evolutionary \cite{francois_predicting_2010} approaches could in principle serve as comparisons, but they lack the throughput needed to sample thousands of networks, and they have their own, different biases. This raises the broader question of the inductive biases of our optimization approach, which might or might not be shared with actual evolution. Such biases could arise, for instance, from combining multiple constraints into a single loss rather than using a Pareto-like optimization~\cite{warmflash_pareto_2012}. A bias toward simpler or more symmetric solutions during optimization has already been reported in related contexts~\cite{johnston2022symmetry, mochulska2025generative}, and has been related to the idea from complexity theory that ``simple'' patterns (in the Kolmogorov complexity sense) are, surprisingly, more probable~\cite{dingle2018input}. Here, symmetric patterns are already common among high-information solutions but come to dominate once scaling is imposed, suggesting that such intrinsic biases interact non-trivially with the geometry of the constrained solution space. These questions could be explored directly within our framework by varying the optimization procedure or its initial conditions. Notably, recent large-scale simulations of artificial neural networks have also shown increasingly complex relationships between initial conditions and convergence time as problems become harder, including a fractal phase transition~\cite{lai2026fractal}. Whether similar phenomena occur for GRNs remains an open question.

Lastly, the coexistence of multiple regimes in our optimizations, with many solutions on the one hand vs few solutions on the other hand is very reminiscent of the evolutionary question raised in the introduction of contingency vs convergence. The size and diversity of network spaces have been long studied through the lens of robustness and evolvability \cite{ciliberti2007robustness, wagner2013robustness, wagner2015arrival} or through enumeration of circuits performing specific tasks \cite{ma2009defining, cotterell2010atlas}. While these approaches were typically limited to small networks and/or discrete phenotypes, our framework complements and extends them: it optimizes a continuous loss (e.g.\ information) through stochastic spatial simulations, can sample thousands of solutions for a given set of constraints, and track how the ensemble of solutions reorganizes. In our simulations, one can further continuously move from one regime to the other simply by adding more constraints. This does not tell us how real evolution behaves, but shows that our proposed framework can turn the old contingency vs convergence debate into a series of testable questions at scale: given a biological function and a set of constraints, how large, connected and diverse is the space of solutions? More broadly, differentiable programming may therefore provide the basis for a \textit{variational systems biology} in which the possible, and not only the actual, becomes an object of systematic study.

\section{Methods}\label{sec:methods}

\subsection{Model and stochastic simulations}\label{sec:model}

We simulate morphogen-driven reaction-diffusion dynamics along a one-dimensional anterior-posterior (AP) axis for an ensemble of $N_e$ embryos, with $N_m$ externally specified morphogen inputs and $N_g$ network genes interacting through regulatory feedback. For each gene $\alpha$,
\begin{equation}
\frac{\partial g_\alpha}{\partial t} = P_\alpha\!\left(\mathbf g,\mathbf c\right) - \lambda_\alpha\, g_\alpha + D_\alpha \frac{\partial^2 g_\alpha}{\partial x^2}
\label{eq:pde_continuous}
\end{equation}
where $\mathbf g$ and $\mathbf c$ are the vectors of gene and morphogen levels, $\lambda_\alpha$ the decay rate, $D_\alpha$ the diffusion coefficient and $P_\alpha$ the production term. We integrate a spatially discretized version of Eq.~\ref{eq:pde_continuous} on $N_x$ nuclear volumes separated by a distance $\Delta$, with reflecting boundaries and effective hopping rate $h_\alpha=D_\alpha/\Delta^2$.

Gene regulation is modeled with a Monod-Wyman-Changeux (MWC) framework rather than a phenomenological Hill function, so that each interaction is described by two interpretable parameters: a signed regulatory strength $H$ and an effective dissociation constant $K$. The activity of gene $\alpha$ is a sigmoid of a regulation potential combining saturating morphogen and network contributions,
\begin{equation}
\begin{aligned}
f_\alpha &= \frac{1}{1+e^{-F_\alpha}},\\
F_\alpha &= \sum_{\kappa=1}^{N_m} H^{M}_{\alpha\kappa} \log\!\left(1+\frac{c_{\kappa,i}}{K^{M}_{\alpha\kappa}}\right)\\
&\quad + \sum_{\zeta=1}^{N_g} H^{G}_{\alpha\zeta} \log\!\left(1+\frac{g_{\zeta,i}}{K^{G}_{\alpha\zeta}}\right) - F^{(0)}_{\alpha}
\end{aligned}
\label{eq:mwc}
\end{equation}
where $F^{(0)}$ sets gene-specific activation thresholds and the $M$/$G$ superscripts distinguish morphogen and network gene contributions. When prior knowledge is available about interaction signs, we optionally fix the sign of selected entries of $H^M$ and/or $H^G$ while learning their magnitudes. The production term includes a basal floor, \mbox{$P_\alpha = P_{\mathrm{basal}} + (1-P_{\mathrm{basal}})f_\alpha$}.

Stochasticity enters at two levels. Embryo-to-embryo extrinsic variability is applied multiplicatively to the inputs: for each embryo $e$, input $\kappa$ and site $i$,
\begin{equation}
\tilde c^{(e)}_{\kappa,i} = c_{\kappa,i} \left(1+\sigma_{\mathrm{amp}}\,\epsilon^{\mathrm{amp}}_{\kappa}\right) \left(1+\sigma_{\mathrm{local}}\,\epsilon^{\mathrm{local}}_{\kappa,i}\right)
\label{eq:input_noise}
\end{equation}
with $\epsilon\sim\mathcal N(0,1)$ generating global amplitude and local fluctuations, drawn once per embryo and held fixed along the trajectory; we write $\sigma_{\mathrm{IN}}$ for their common value when $\sigma_{\mathrm{amp}}=\sigma_{\mathrm{local}}$, calibrated against measured Bicoid variability (Supplement). Intrinsic expression noise is added during integration through a chemical Langevin approximation with production, degradation and diffusive-hopping contributions~\cite{gillespie_chemical_2000,de_jong_modeling_2002,sokolowski_deriving_2025}. Equations are integrated by explicit Euler-Maruyama with a ReLU projection enforcing non-negative expression; differentiability is preserved with a straight-through estimator (Supplement).

Morphogens have fixed physical decay lengths $\ell$ (reported in units of the reference length $L_0$), so that in relative coordinates $x_s=x/L$ their profiles
\begin{equation}
c(x_s) = \exp\!\left(-L\,x_s/\ell\right), \qquad 0<x_s<1
\label{eq:non_scaling_gradient}
\end{equation}
depend explicitly on embryo length and are not scale invariant; terminal inputs expressed at both poles are built as the sum of two such gradients (Supplement).

Embryo lengths are drawn as $L_e=L_0(1+\sigma_L\xi_e)$, $\xi_e\sim\mathcal N(0,1)$. All embryos share the same grid of relative positions, so a length change rescales the spacing, $\Delta=s\Delta_0$ with $s=L_e/L_0$: size therefore enters both through the input profiles (Eq.~\ref{eq:non_scaling_gradient}) and through the effective hopping rate $h_\alpha=D_\alpha/(s\Delta_0)^2$. Although the framework supports optimization of dynamical parameters, diffusion and decay are here held fixed and identical across genes, so that $\theta=\{H^M,H^G,K^M,K^G,F^{(0)}\}$ (bounded reparameterizations, Supplement). Noise sources and embryo lengths are resampled at every optimization step, and all settings are listed in Table~\ref{tab:settings}. Following~\cite{sokolowski_deriving_2025}, the posteriormost $7\%$ of embryo length is excluded from objectives and position averages in the positional-information problem, as wild-type gap genes are not expressed there.

\begin{table*}[tp]
\centering
\caption{Simulation and optimization settings.}
\label{tab:settings}
\begin{tabular}{@{}lll@{}}
\toprule
Nuclei & $N_x$ & $90$~\cite{mcgough_finding_2024} \\
Reference spacing & $\Delta_0$ & $8.5~\mathrm{\mu m}$~\cite{sokolowski_deriving_2025} \\
Length variability & $\sigma_L$ & $5\%$~\cite{nikolic_scale_2024} \\
Diffusion & $D_\alpha$ & $0.4~\mathrm{\mu m^2/s}$~\cite{sokolowski_deriving_2025} \\
Decay rate & $\lambda_\alpha$ & $1$ (sets the time unit, $\tau=\lambda_{\mathrm{phys}}^{-1}=10$~min) \\
Basal production & $P_{\mathrm{basal}}$ & $10^{-3}$ \\
Intrinsic noise & $\eta$ & $3\times10^{-3}$ \\
Input noise & $\sigma_{\mathrm{IN}}$ & $0$ (domain formation) / $[0, 0.05, 0.10, 0.15]$ (positional information) \\
Embryos per step & $N_e$ & $100$ \\
Time step & $dt$ & $0.01$ \\
Measurement time & $t_{\mathrm{meas}}$ & $15$ decay times \\
Optimizer & Adam~\cite{kingma_adam_2017} & learning rate $0.1$, $5000$ steps \\
Runs per scenario & & $200$ random initializations \\
Analysis ensemble & & $500$ embryos \\
\bottomrule
\end{tabular}
\end{table*}

\subsection{Loss structure}\label{sec:loss}

We minimize by gradient descent a loss combining a primary objective with biological constraints and regularization,
\begin{equation}
\mathcal{L}_{\mathrm{total}}(\theta) = \mathcal{L}_{\mathrm{p}}\!\left(\mathbf g_\theta\right) + \sum_{c} \lambda_c \,\mathcal{L}_c\!\left(\mathbf g_\theta\right) + \lambda_{\mathrm{reg}} \,\mathcal{L}_{\mathrm{reg}}(\theta)
\label{eq:loss_simple}
\end{equation}
evaluated on the simulated profiles $\mathbf g_\theta=\{g^{(e)}_{\alpha,i}(t_n)\}$, where $\mathcal{L}_{\mathrm{p}}$ is the primary objective, either penalizing an error directly or written with a minus sign to maximize a metric (e.g.\ positional information, $\mathcal{L}_{\mathrm p}=-I$). Constraints are soft inequality penalties~\cite{nocedal_2006_numerical}, inactive when a model output falls within a biologically motivated range and growing smoothly once it is violated: for a quantity $Q$ with admissible threshold $Q_{\mathrm{target}}$,
\begin{equation}
\mathcal{L}_{Q} = p_Q^{\,2}, \qquad p_Q = \frac{\max(0,\, Q - Q_{\mathrm{target}})}{Q_{\mathrm{target}}}
\label{eq:loss_constraints}
\end{equation}
with symmetric penalties for two-sided ranges. Normalization by $Q_{\mathrm{target}}$ renders constraints dimensionless and comparable. Two constraints are used in the positional-information problem: a global resource-utilization budget, defined as expression averaged over genes and spatial positions at $t_{\mathrm{meas}}$, with an upper bound of $RU_{\max}=0.25$ as measured in gap gene data~\cite{petkova_optimal_2019}, and a minimum peak-expression target of $0.9$ per gene. The regularization term $\mathcal{L}_{\mathrm{reg}}$ consists of an $L_2$ penalty on regulatory parameters, supplemented in the domain-formation problem by a cosine-similarity penalty on near-duplicate rows of $H^M$ and $H^G$ to discourage different genes from converging to identical regulatory structures (Supplement). A constraint may alternatively be applied as a multiplicative gate on the primary objective rather than as an additive penalty; we use this only for the scaling constraint, described below.

\subsection{Loss function for domain formation}\label{sec:domain_loss}

Here position is a discrete variable $X$ with values $i\in\{1,\dots,N_x\}$ and cell-fate identity a discrete variable $Y$ with values $\alpha\in\{1,\dots,N_g\}$. Following~\cite{francois_predicting_2010}, we define per embryo the transformed expression $\tilde g^{(e)}_{\alpha,i}=g^{(e)}_{\alpha,i}/(1+g^{(e)}_{\alpha,i})$, which limits the dominance of highly expressed genes, and convert it into a categorical distribution over identities,
\begin{equation}
P^{(e)}(\alpha\mid i) = \tilde g^{(e)}_{\alpha,i}\Big/\sum\nolimits_{\beta=1}^{N_g}\tilde g^{(e)}_{\beta,i}
\label{eq:categorical_identity}
\end{equation}
so that identity depends on relative rather than absolute expression. These conditionals are averaged across embryos at each relative position, $P(\alpha\mid i)=\frac{1}{N_e}\sum_e P^{(e)}(\alpha\mid i)$, and combined with a uniform prior $P(i)=1/N_x$ into the joint $P(i,\alpha)$. The domain-formation loss is the logarithmic distance from the maximal achievable position-identity information $\log N_g$~\cite{francois_predicting_2010},
\begin{equation}
\begin{aligned}
\mathcal L_{\mathrm{domain}} &= \log\!\left(\log N_g-I(X;Y)\right),\\
I(X;Y) &= \sum_{i,\alpha}P(i,\alpha)\log\!\frac{P(i,\alpha)}{P(i)P(\alpha)}
\end{aligned}
\label{eq:loss_domain}
\end{equation}
where $I(X;Y)$ reaches its maximum $\log N_g$ for $N_g$ sharply separated domains of equal extent and vanishes when identities are uniformly mixed, so that minimizing $\mathcal L_{\mathrm{domain}}$ drives $I(X;Y)$ toward its maximum.

\subsection{Differentiable positional information}\label{sec:pi}

Positional information is the mutual information between the expression vector $\mathbf g$, measured at a fixed readout time, and position $x$:
\begin{equation}
I(\mathbf g;x) = S[P(\mathbf g)] - \big\langle S[P(\mathbf g\mid x)] \big\rangle_x
\label{eq:pi_def}
\end{equation}
where $S[\cdot]$ denotes entropy and spatial averages are discrete sums. Throughout, $I$ denotes this quantity, expressed in bits, and is evaluated for all optimized models irrespective of the objective used. Following~\cite{dubuis_positional_2013, sokolowski_deriving_2025}, the conditional distribution at fixed position is approximated by a multivariate Gaussian $\mathcal N(\mathbf g;\bar{\mathbf g}(x),C(x))$ with $\bar{\mathbf g}$ and $C$ the ensemble mean and covariance, giving $S[P(\mathbf g\mid x)]=\tfrac12\log[(2\pi e)^{N_g}\det C(x)]$. The marginal $P(\mathbf g)=\sum_i P(x_i)P(\mathbf g\mid x_i)$ is then a Gaussian mixture whose entropy has no closed form, as it involves the logarithm of a sum of Gaussians. We approximate it by a second-order expansion~\cite{huber_entropy_2008}, Taylor expanding $\log P(\mathbf g)$ about the conditional mean of each mixture component and averaging under the local Gaussian:
\begin{equation}
\begin{aligned}
S[P(\mathbf g)] \approx -\bigg\langle &\log P(\bar{\mathbf g}(x)) \\
&+ \frac{1}{2}\, \mathrm{tr}\!\left(C(x)\, \nabla^2_{\mathbf g}\log P(\mathbf g) \Big|_{\mathbf g=\bar{\mathbf g}(x)} \right) \bigg\rangle_x
\end{aligned}
\label{eq:huber_entropy}
\end{equation}
where the second term is a curvature correction controlled by the local covariance. Eq.~\ref{eq:huber_entropy} yields a differentiable approximation of $I(\mathbf g;x)$ computed entirely from simulated ensemble means and covariances, through which gradients are propagated together with the simulation pipeline itself. Precision is quantified by the positional error~\cite{dubuis_positional_2013}
\begin{equation}
\sigma_x(x)^{-2} = \sum_{\alpha,\beta=1}^{N_g} \frac{d\bar g_{\alpha}}{dx} \left(C^{-1}(x)\right)_{\alpha\beta} \frac{d\bar g_{\beta}}{dx}
\label{eq:pos_error}
\end{equation}
with derivatives taken by central differences in relative coordinates and $\sigma_x$ reported in cells, one cell being one nuclear spacing. The positional-error uniformity constraint minimizes the coefficient of variation $\mathrm{CV}(\sigma_x)$ along the axis, which has no natural target value.

\subsection{Scale invariance}\label{sec:scaling}

Scale invariance is optimized explicitly through a constraint on the mean absolute deviation (MAD) of expression profiles, which measures profile alignment in relative coordinates as the simulated embryonic length varies,
\begin{equation}
M = \frac{1}{N_x N_g} \sum_{i=1}^{N_x} \sum_{\alpha=1}^{N_g} \frac{1}{N_e} \sum_{e=1}^{N_e} \left| g^{(e)}_{\alpha,i} - \bar g_{\alpha,i} \right|
\label{eq:L_scale}
\end{equation}
with $\bar g_{\alpha,i}$ the ensemble mean profile (excluding the posterior cutoff region in the positional-information problem). The corresponding penalty $\mathcal L_{\mathrm{MAD}}$ is given by Eq.~\ref{eq:loss_constraints} with $Q=M$ and $Q_{\mathrm{target}}=0.01$. It is applied as a multiplicative reward gate on the primary objective~\cite{warmflash_pareto_2012, sokolowski_deriving_2025},
\begin{equation}
\begin{aligned}
&\mathcal L_{\mathrm p} \longrightarrow
\begin{cases}
G\,\mathcal L_{\mathrm p}, & \mathcal L_{\mathrm p}<0,\\
\mathcal L_{\mathrm p}/G, & \mathcal L_{\mathrm p}\geq 0,
\end{cases}\\
&G = 1+e^{-\beta\,\mathcal L_{\mathrm{MAD}}}, \quad \beta=0.05,
\end{aligned}
\label{eq:loss_gated}
\end{equation}
so that the gate $G$ equals two when the constraint is satisfied and decays toward one as it is violated, thereby favoring alignment. In the positional-information problem, $\mathcal L_{\mathrm p}=-I$ is negative, so the gate simply becomes $-I \longrightarrow -I\,G$. In the domain-formation problem, the gate acts on the logarithmic loss of Eq.~\ref{eq:loss_domain}.

After optimization, boundaries are detected on individual gene expression profiles (Supplement), and expressed in relative coordinates $x_b/L$, which are length-independent under perfect scaling (Fig.~\ref{fig:MI_scaling_example}C). We compute their embryo-to-embryo standard deviation $\sigma_{x_b}$, compared to the value expected for non-scaling boundaries anchored to the poles (Supplement), and the systematic boundary shift $\Delta x_b$, obtained by linear regression of $x_b/L$ against embryo length and expressed in cells over the $5$--$95\%$ range of simulated lengths. Following~\cite{nikolic_scale_2024}, we also quantify the deviation from scale invariance
\begin{equation}
\Delta I = I(\mathbf g; \{x, L\}) - I(\mathbf g; x/L)
\label{eq:deltaI}
\end{equation}
which measures how much positional information is gained by additionally knowing embryo length, so that $\Delta I\simeq0$ means length cannot be inferred from the profiles, a necessary condition for scaling. It is estimated under a Gaussian approximation on embryos binned by length and corrected for finite-sample bias by extrapolating subsampled estimates to infinitely many embryos, with bootstrapped errors and shuffled controls (example in Fig.~\ref{fig:MI_scaling_example}E; Supplement). For the domain-formation problem, a GRN is classified as fully scaling if all expressed genes satisfy $\Delta I<0.02$ bits, $\Delta I/I<2\%$ and $\Delta x_b<1$ cell. For the positional-information problem, we use slightly relaxed thresholds to account for the added input noise: all four genes must satisfy $\Delta I<0.05$ bits, $\Delta I/I<2\%$ and $\Delta x_b<2$ cells. These thresholds remain consistent with the range of scaling metrics measured experimentally in the \textit{Drosophila} gap gene system~\cite{nikolic_scale_2024}. A GRN is classified as partially scaling if at least one, but not all, of the genes satisfies the corresponding criteria. Runs are classified as not converged when the loss plateaus well above the ensemble mean loss, and as forming no pattern when the profiles carry less than $3$ bits of positional information.

\subsection{Network analysis}\label{sec:network}

Following~\cite{sokolowski_deriving_2025}, each interaction between regulator $i$ (gene or morphogen) and target gene $j$ is summarized by its effective contribution at maximal expression, \mbox{$W_{ij}=H_{ij}\log(1+1/K_{ij})$}, classified as negligible ($|W_{ij}|<\theta_w$), weak ($\theta_w\le|W_{ij}|<\theta_s$) or strong ($|W_{ij}|\ge\theta_s$), with $\theta_w=1$ and $\theta_s=10$; signed-score matrices summarizing sign consistency across an ensemble are defined in the Supplement. Regulatory topologies are compared by discretizing each $W_{ij}$ into five signed categories using the same thresholds, reordering genes by first half-maximum crossing and concatenating the discretized input-to-gene and gene-to-gene matrices into one topology vector per GRN, which are then clustered by Manhattan distance and average linkage (Supplement). Optimized patterns are compared across runs through a permutation-invariant $L^p$ distance between mean profiles at $t_{\mathrm{meas}}$,
\begin{equation}
D_p(\mathbf g,\hat{\mathbf g}) = \min_{\pi} \left[\frac{1}{N_g N_x} \sum_{\alpha=1}^{N_g} \sum_{i=1}^{N_x} \left| \bar g_{\alpha,i}-\hat g_{\pi(\alpha),i} \right|^p \right]^{1/p}
\label{eq:pattern_distance}
\end{equation}
minimized over permutations $\pi$ of the gene labels, which are arbitrary across independent optimizations; we set $p=3$ following~\cite{sokolowski_deriving_2025} and reorder $D_3$ matrices by hierarchical clustering (average linkage). Finally, optimized networks are perturbed by single-gene knockouts and by setting selected classes of entries of $H^G$ to zero: removing auto-regulation, nearest-neighbor (NN) interactions or all gene--gene interactions, or retaining only auto-regulation, only NN interactions, or both. The last condition removes all non-adjacent interactions, i.e.\ both next-nearest-neighbor (NNN) and longer-range interactions, with neighbors defined by ordering genes by first half-maximum crossing. Perturbed networks are then re-simulated without re-optimization and re-analyzed with the same metrics (Supplement).

\section*{Code availability}
The \texttt{jaxgrn} framework and the configuration files used to generate all figures will be released as open-source software upon journal publication. A preliminary version is available from the authors on request.

\section*{Acknowledgments}
We acknowledge the use of AI-based writing tools for light language enhancement. This work was supported by Institut Pasteur, the French National Research Agency (ANR-10-LABX-73 ``Revive'' and ANR-16-CONV-0005 ``INCEPTION''), the CFM Foundation for Research, and the inIdEx FORMULA project at Universit\'e Paris Cit\'e (ANR-18-IDEX-0001). This research was enabled in part by support provided by Calcul Québec (\url{calculquebec.ca}) and the Digital Research Alliance of Canada (\url{alliancecan.ca}).

\bibliographystyle{ieeetr}
\bibliography{references}

@article{mangan2003structure,
  title   = {Structure and function of the feed-forward loop network motif},
  author  = {Mangan, Shmoolik and Alon, Uri},
  journal = {Proceedings of the National Academy of Sciences},
  volume  = {100},
  number  = {21},
  pages   = {11980--11985},
  year    = {2003}
}

@article{nilsson1994pessimistic,
  title   = {A pessimistic estimate of the time required for an eye to evolve},
  author  = {Nilsson, Dan-E. and Pelger, Susanne},
  journal = {Proceedings of the Royal Society B: Biological Sciences},
  volume  = {256},
  number  = {1345},
  pages   = {53--58},
  year    = {1994}
}

@article{jacob1977evolution,
  title   = {Evolution and tinkering},
  author  = {Jacob, Fran{\c{c}}ois},
  journal = {Science},
  volume  = {196},
  number  = {4295},
  pages   = {1161--1166},
  year    = {1977}
}

@article{cotterell2010atlas,
  title   = {An atlas of gene regulatory networks reveals multiple three-gene mechanisms for interpreting morphogen gradients},
  author  = {Cotterell, James and Sharpe, James},
  journal = {Molecular Systems Biology},
  volume  = {6},
  pages   = {425},
  year    = {2010}
}

@book{bialek2012biophysics,
  title     = {Biophysics: Searching for Principles},
  author    = {Bialek, William},
  publisher = {Princeton University Press},
  year      = {2012}
}

@article{franccois2007deriving,
  title   = {Deriving structure from evolution: metazoan segmentation},
  author  = {Fran{\c{c}}ois, Paul and Hakim, Vincent and Siggia, Eric D.},
  journal = {Molecular Systems Biology},
  volume  = {3},
  pages   = {154},
  year    = {2007}
}

@article{lai2026fractal,
  title   = {Fractal basins trap latent reasoning},
  author  = {Lai, Jeffrey and Bao, Anthony and Quinn, John and Gilpin, William},
  journal = {arXiv preprint arXiv:2609.04963},
  year    = {2026}
}

@article{booth2025gene,
  title   = {Gene network organization, mutation, and selection collectively drive developmental pattern evolvability and predictability},
  author  = {Booth, Harry and Hadjivasiliou, Zena},
  journal = {PRX Life},
  volume  = {3},
  number  = {3},
  pages   = {033023},
  year    = {2025}
}

@article{lalanne2013principles,
  title   = {Principles of adaptive sorting revealed by in silico evolution},
  author  = {Lalanne, Jean-Beno{\^\i}t and Fran{\c{c}}ois, Paul},
  journal = {Physical Review Letters},
  volume  = {110},
  number  = {21},
  pages   = {218102},
  year    = {2013}
}

@article{kondo2025engineering,
  title   = {Engineering {TCR}-controlled fuzzy logic into {CAR T} cells enhances therapeutic specificity},
  author  = {Kondo, Taisuke and Bourassa, Fran{\c{c}}ois X. P. and Achar, Sooraj and DuSold, Justyn and C{\'e}spedes, Pablo F. and Ando, Makoto and Dwivedi, Alka and Moraly, Josquin and Chien, Christopher and Majdoul, Saliha and others},
  journal = {Cell},
  volume  = {188},
  number  = {9},
  pages   = {2372--2389},
  year    = {2025}
}

@book{wagner2015arrival,
  title     = {Arrival of the Fittest: How Nature Innovates},
  author    = {Wagner, Andreas},
  publisher = {Current},
  year      = {2015}
}

@book{erwin2026origins,
  title     = {The Origins of the New: Novelty and Innovation in the History of Life, Culture, and Technology},
  author    = {Erwin, Douglas H.},
  publisher = {Princeton University Press},
  year      = {2026}
}

@book{gould1989wonderful,
  title     = {Wonderful Life: The Burgess Shale and the Nature of History},
  author    = {Gould, Stephen Jay},
  publisher = {W. W. Norton \& Company},
  year      = {1989}
}

@article{losos2011convergence,
  title   = {Convergence, adaptation, and constraint},
  author  = {Losos, Jonathan B.},
  journal = {Evolution},
  volume  = {65},
  number  = {7},
  pages   = {1827--1840},
  year    = {2011}
}

@article{uller2018developmental,
  title   = {Developmental bias and evolution: A regulatory network perspective},
  author  = {Uller, Tobias and Moczek, Armin P. and Watson, Richard A. and Brakefield, Paul M. and Laland, Kevin N.},
  journal = {Genetics},
  volume  = {209},
  number  = {4},
  pages   = {949--966},
  year    = {2018}
}

@article{kaneko2007evolution,
  title   = {Evolution of robustness to noise and mutation in gene expression dynamics},
  author  = {Kaneko, Kunihiko},
  journal = {{PLoS ONE}},
  volume  = {2},
  number  = {5},
  pages   = {e434},
  year    = {2007}
}

@article{furusawa2018formation,
  title   = {Formation of dominant mode by evolution in biological systems},
  author  = {Furusawa, Chikara and Kaneko, Kunihiko},
  journal = {Physical Review E},
  volume  = {97},
  number  = {4},
  pages   = {042410},
  year    = {2018}
}

@article{husain2020physical,
  title   = {Physical constraints on epistasis},
  author  = {Husain, Kabir and Murugan, Arvind},
  journal = {Molecular Biology and Evolution},
  volume  = {37},
  number  = {10},
  pages   = {2865--2874},
  year    = {2020}
}

@article{schaerli2018synthetic,
  title   = {Synthetic circuits reveal how mechanisms of gene regulatory networks constrain evolution},
  author  = {Schaerli, Yolanda and Jim{\'e}nez, Alba and Duarte, Jos{\'e} M. and Mihajlovic, Ljiljana and Renggli, Julien and Isalan, Mark and Sharpe, James and Wagner, Andreas},
  journal = {Molecular Systems Biology},
  volume  = {14},
  number  = {9},
  pages   = {MSB178102},
  year    = {2018}
}

@article{schaerli2014unified,
  title   = {A unified design space of synthetic stripe-forming networks},
  author  = {Schaerli, Yolanda and Munteanu, Andreea and Gili, Mag{\"u}i and Cotterell, James and Sharpe, James and Isalan, Mark},
  journal = {Nature Communications},
  volume  = {5},
  number  = {1},
  pages   = {4905},
  year    = {2014}
}

@article{kaplan2020scaling,
  title   = {Scaling laws for neural language models},
  author  = {Kaplan, Jared and McCandlish, Sam and Henighan, Tom and Brown, Tom B. and Chess, Benjamin and Child, Rewon and Gray, Scott and Radford, Alec and Wu, Jeffrey and Amodei, Dario},
  journal = {arXiv preprint arXiv:2001.08361},
  year    = {2020}
}

@article{gregor2005diffusion,
  title   = {Diffusion and scaling during early embryonic pattern formation},
  author  = {Gregor, Thomas and Bialek, William and {de Ruyter van Steveninck}, Rob R. and Tank, David W. and Wieschaus, Eric F.},
  journal = {Proceedings of the National Academy of Sciences},
  volume  = {102},
  number  = {51},
  pages   = {18403--18407},
  year    = {2005}
}

@article{ben2008scaling,
  title   = {Scaling of the {BMP} activation gradient in {Xenopus} embryos},
  author  = {Ben-Zvi, Danny and Shilo, Ben-Zion and Fainsod, Abraham and Barkai, Naama},
  journal = {Nature},
  volume  = {453},
  number  = {7199},
  pages   = {1205--1211},
  year    = {2008}
}

@article{lauschke2013scaling,
  title   = {Scaling of embryonic patterning based on phase-gradient encoding},
  author  = {Lauschke, Volker M. and Tsiairis, Charisios D. and Fran{\c{c}}ois, Paul and Aulehla, Alexander},
  journal = {Nature},
  volume  = {493},
  number  = {7430},
  pages   = {101--105},
  year    = {2013}
}

@article{merle2024precise,
  title   = {Precise and scalable self-organization in mammalian pseudo-embryos},
  author  = {Merle, Melody and Friedman, Leah and Chureau, Corinne and Shoushtarizadeh, Armin and Gregor, Thomas},
  journal = {Nature Structural \& Molecular Biology},
  volume  = {31},
  number  = {6},
  pages   = {896--902},
  year    = {2024}
}

@article{cooke1981scale,
  title   = {Scale of body pattern adjusts to available cell number in amphibian embryos},
  author  = {Cooke, Jonathan},
  journal = {Nature},
  volume  = {290},
  number  = {5809},
  pages   = {775--778},
  year    = {1981}
}

@article{wilke2001evolution,
  title   = {Evolution of digital organisms at high mutation rates leads to survival of the flattest},
  author  = {Wilke, Claus O. and Wang, Jia Lan and Ofria, Charles and Lenski, Richard E. and Adami, Christoph},
  journal = {Nature},
  volume  = {412},
  number  = {6844},
  pages   = {331--333},
  year    = {2001}
}

@article{johnston2022symmetry,
  title   = {Symmetry and simplicity spontaneously emerge from the algorithmic nature of evolution},
  author  = {Johnston, Iain G. and Dingle, Kamaludin and Greenbury, Sam F. and Camargo, Chico Q. and Doye, Jonathan P. K. and Ahnert, Sebastian E. and Louis, Ard A.},
  journal = {Proceedings of the National Academy of Sciences},
  volume  = {119},
  number  = {11},
  pages   = {e2113883119},
  year    = {2022}
}

@article{mochulska2025generative,
  title   = {Generative epigenetic landscapes map the topology and topography of cell fates},
  author  = {Mochulska, Victoria and Fran{\c{c}}ois, Paul},
  journal = {Proceedings of the National Academy of Sciences},
  volume  = {122},
  number  = {50},
  pages   = {e2514508122},
  year    = {2025}
}

@article{dingle2018input,
  title   = {Input--output maps are strongly biased towards simple outputs},
  author  = {Dingle, Kamaludin and Camargo, Chico Q. and Louis, Ard A.},
  journal = {Nature Communications},
  volume  = {9},
  number  = {1},
  pages   = {761},
  year    = {2018}
}

@article{sokolowski_deriving_2025,
  title   = {Deriving a genetic regulatory network from an optimization principle},
  author  = {Sokolowski, Thomas R. and Gregor, Thomas and Bialek, William and Tka{\v{c}}ik, Ga{\v{s}}per},
  journal = {Proceedings of the National Academy of Sciences},
  volume  = {122},
  number  = {1},
  pages   = {e2402925121},
  year    = {2025}
}

@article{gillespie_chemical_2000,
  title   = {The chemical {Langevin} equation},
  author  = {Gillespie, Daniel T.},
  journal = {The Journal of Chemical Physics},
  volume  = {113},
  number  = {1},
  pages   = {297--306},
  year    = {2000}
}

@book{kloeden_numerical_1992,
  title     = {Numerical Solution of Stochastic Differential Equations},
  author    = {Kloeden, Peter E. and Platen, Eckhard},
  publisher = {Springer},
  address   = {Berlin, Heidelberg},
  year      = {1992}
}

@article{alquraishi_differentiable_2021,
  title   = {Differentiable biology: using deep learning for biophysics-based and data-driven modeling of molecular mechanisms},
  author  = {{AlQuraishi}, Mohammed and Sorger, Peter K.},
  journal = {Nature Methods},
  volume  = {18},
  number  = {10},
  pages   = {1169--1180},
  year    = {2021}
}

@article{francois_predicting_2010,
  title   = {Predicting embryonic patterning using mutual entropy fitness and in silico evolution},
  author  = {Fran{\c{c}}ois, Paul and Siggia, Eric D.},
  journal = {Development},
  volume  = {137},
  number  = {14},
  pages   = {2385--2395},
  year    = {2010}
}

@article{walczak_optimizing_2010,
  title   = {Optimizing information flow in small genetic networks. {II}. Feed-forward interactions},
  author  = {Walczak, Aleksandra M. and Tka{\v{c}}ik, Ga{\v{s}}per and Bialek, William},
  journal = {Physical Review E},
  volume  = {81},
  number  = {4},
  pages   = {041905},
  year    = {2010}
}

@article{deistler_jaxley_2025,
  title   = {{Jaxley}: differentiable simulation enables large-scale training of detailed biophysical models of neural dynamics},
  author  = {Deistler, Michael and Kadhim, Kyra L. and Pals, Matthijs and Beck, Jonas and Huang, Ziwei and Gloeckler, Manuel and Lappalainen, Janne K. and Schr{\"o}der, Cornelius and Berens, Philipp and Gon{\c{c}}alves, Pedro J. and Macke, Jakob H.},
  journal = {Nature Methods},
  volume  = {22},
  number  = {12},
  pages   = {2649--2657},
  year    = {2025}
}

@article{de_jong_modeling_2002,
  title   = {Modeling and simulation of genetic regulatory systems: a literature review},
  author  = {de Jong, Hidde},
  journal = {Journal of Computational Biology},
  volume  = {9},
  number  = {1},
  pages   = {67--103},
  year    = {2002},
  doi     = {10.1089/10665270252833208}
}

@article{jaeger_dynamical_2004,
  title   = {Dynamical analysis of regulatory interactions in the gap gene system of {Drosophila melanogaster}},
  author  = {Jaeger, Johannes and Blagov, Maxim and Kosman, David and Kozlov, Konstantin N. and Manu and Myasnikova, Ekaterina and Surkova, Svetlana and Vanario-Alonso, Carlos E. and Samsonova, Maria and Sharp, David H. and Reinitz, John},
  journal = {Genetics},
  volume  = {167},
  number  = {4},
  pages   = {1721--1737},
  year    = {2004},
  doi     = {10.1534/genetics.104.027334}
}

@article{tkacik_information_2008,
  title   = {Information flow and optimization in transcriptional regulation},
  author  = {Tka{\v{c}}ik, Ga{\v{s}}per and Callan, Curtis G. and Bialek, William},
  journal = {Proceedings of the National Academy of Sciences},
  volume  = {105},
  number  = {34},
  pages   = {12265--12270},
  year    = {2008},
  doi     = {10.1073/pnas.0806077105}
}

@article{francois_design_2004,
  title   = {Design of genetic networks with specified functions by evolution in silico},
  author  = {Fran{\c{c}}ois, Paul and Hakim, Vincent},
  journal = {Proceedings of the National Academy of Sciences},
  volume  = {101},
  number  = {2},
  pages   = {580--585},
  year    = {2004},
  doi     = {10.1073/pnas.0304532101}
}

@article{dubuis_positional_2013,
  title   = {Positional information, in bits},
  author  = {Dubuis, Julien O. and Tka{\v{c}}ik, Ga{\v{s}}per and Wieschaus, Eric F. and Gregor, Thomas and Bialek, William},
  journal = {Proceedings of the National Academy of Sciences},
  volume  = {110},
  number  = {41},
  pages   = {16301--16308},
  year    = {2013},
  doi     = {10.1073/pnas.1315642110}
}

@article{sokolowski_mutual_2012,
  title   = {Mutual repression enhances the steepness and precision of gene expression boundaries},
  author  = {Sokolowski, Thomas R. and Erdmann, Thorsten and {ten Wolde}, Pieter Rein},
  journal = {{PLOS} Computational Biology},
  volume  = {8},
  number  = {8},
  pages   = {e1002654},
  year    = {2012},
  doi     = {10.1371/journal.pcbi.1002654}
}

@article{huber_entropy_2008,
  title   = {On entropy approximation for {Gaussian} mixture random vectors},
  author  = {Huber, Marco F. and Bailey, Tim and Durrant-Whyte, Hugh and Hanebeck, Uwe D.},
  journal = {Proceedings of the IEEE International Conference on Multisensor Fusion and Integration for Intelligent Systems},
  pages   = {181--188},
  year    = {2008},
  doi     = {10.1109/MFI.2008.4648062}
}

@article{mcgough_finding_2024,
  title   = {Finding the last bits of positional information},
  author  = {{McGough}, Lauren and Casademunt, Helena and Nikoli{\'c}, Milo{\v{s}} and Aridor, Zoe and Petkova, Mariela D. and Gregor, Thomas and Bialek, William},
  journal = {{PRX} Life},
  volume  = {2},
  number  = {1},
  pages   = {013016},
  year    = {2024},
  doi     = {10.1103/PRXLife.2.013016}
}

@article{petkova_optimal_2019,
  title   = {Optimal decoding of cellular identities in a genetic network},
  author  = {Petkova, Mariela D. and Tka{\v{c}}ik, Ga{\v{s}}per and Bialek, William and Wieschaus, Eric F. and Gregor, Thomas},
  journal = {Cell},
  volume  = {176},
  number  = {4},
  pages   = {844--855.e15},
  year    = {2019},
  doi     = {10.1016/j.cell.2019.01.007}
}

@article{vakulenko_size_2009,
  title   = {Size regulation in the segmentation of {Drosophila}: Interacting interfaces between localized domains of gene expression ensure robust spatial patterning},
  author  = {Vakulenko, Sergei and {Manu} and Reinitz, John and Radulescu, Ovidiu},
  journal = {Physical Review Letters},
  volume  = {103},
  number  = {16},
  pages   = {168102},
  year    = {2009},
  doi     = {10.1103/PhysRevLett.103.168102}
}

@article{shen_scaling_2022,
  title   = {Scaling dictates the decoder structure},
  author  = {Shen, Jingxiang and Liu, Feng and Tang, Chao},
  journal = {Science Bulletin},
  volume  = {67},
  number  = {14},
  pages   = {1486--1495},
  year    = {2022},
  doi     = {10.1016/j.scib.2022.06.014}
}

@article{shen2021finding,
  title   = {Finding gene network topologies for given biological function with recurrent neural network},
  author  = {Shen, Jingxiang and Liu, Feng and Tu, Yuhai and Tang, Chao},
  journal = {Nature Communications},
  volume  = {12},
  number  = {1},
  pages   = {3125},
  year    = {2021}
}

@article{hiscock2019adapting,
  title   = {Adapting machine-learning algorithms to design gene circuits},
  author  = {Hiscock, Tom W.},
  journal = {BMC Bioinformatics},
  volume  = {20},
  number  = {1},
  pages   = {214},
  year    = {2019}
}

@article{nikolic_scale_2024,
  title   = {Scale invariance in early embryonic development},
  author  = {Nikoli{\'c}, Milo{\v{s}} and Antonetti, Victoria and Liu, Feng and Muhaxheri, Gentian and Petkova, Mariela D. and Scheeler, Martin and Smith, Eric M. and Bialek, William and Gregor, Thomas},
  journal = {Proceedings of the National Academy of Sciences},
  volume  = {121},
  number  = {46},
  pages   = {e2403265121},
  year    = {2024},
  doi     = {10.1073/pnas.2403265121}
}

@article{mchale_embryonic_2006,
  title   = {Embryonic pattern scaling achieved by oppositely directed morphogen gradients},
  author  = {{McHale}, Peter and Rappel, Wouter-Jan and Levine, Herbert},
  journal = {Physical Biology},
  volume  = {3},
  number  = {2},
  pages   = {107},
  year    = {2006},
  doi     = {10.1088/1478-3975/3/2/003}
}

@article{warmflash_pareto_2012,
  title   = {Pareto evolution of gene networks: an algorithm to optimize multiple fitness objectives},
  author  = {Warmflash, Aryeh and Fran{\c{c}}ois, Paul and Siggia, Eric D.},
  journal = {Physical Biology},
  volume  = {9},
  number  = {5},
  pages   = {056001},
  year    = {2012},
  doi     = {10.1088/1478-3975/9/5/056001}
}

@article{deshpande_engineering_2025,
  title   = {Engineering morphogenesis of cell clusters with differentiable programming},
  author  = {Deshpande, Ramya and Mottes, Francesco and Vlad, Ariana-Dalia and Brenner, Michael P. and Dal Co, Alma},
  journal = {Nature Computational Science},
  volume  = {5},
  number  = {10},
  pages   = {875--883},
  year    = {2025},
  doi     = {10.1038/s43588-025-00851-4}
}

@article{bengio_estimating_2013,
  title   = {Estimating or propagating gradients through stochastic neurons for conditional computation},
  author  = {Bengio, Yoshua and L{\'e}onard, Nicholas and Courville, Aaron},
  journal = {arXiv preprint arXiv:1308.3432},
  year    = {2013}
}

@inproceedings{kingma_adam_2017,
  title     = {{Adam}: A method for stochastic optimization},
  author    = {Kingma, Diederik P. and Ba, Jimmy},
  booktitle = {International Conference on Learning Representations (ICLR)},
  year      = {2015}
}

@misc{bradbury_jax_2018,
  title  = {{JAX}: Composable transformations of {Python}+{NumPy} programs},
  author = {Bradbury, James and Frostig, Roy and Hawkins, Peter and Johnson, Matthew James and Leary, Chris and Maclaurin, Dougal and Necula, George and Paszke, Adam and VanderPlas, Jake and Wanderman-Milne, Skye and Zhang, Qiao},
  year   = {2018},
  note   = {Version 0.3.13}
}

@book{nocedal_2006_numerical,
  title     = {Numerical Optimization},
  author    = {Nocedal, Jorge and Wright, Stephen J.},
  publisher = {Springer},
  address   = {New York},
  edition   = {2nd},
  year      = {2006}
}

@article{bruckner_marr_2026,
  title   = {Marr's three levels for embryonic development: information, dynamical systems, gene networks},
  author  = {Br{\"u}ckner, David B. and Tka{\v{c}}ik, Ga{\v{s}}per},
  journal = {PRX Life},
  volume  = {4},
  number  = {1},
  pages   = {017001},
  year    = {2026}
}

@article{mlynarski_statistical_2021,
  title   = {Statistical analysis and optimality of neural systems},
  author  = {M{\l}ynarski, Wiktor F. and Hled{\'\i}k, Michal and Sokolowski, Thomas R. and Tka{\v{c}}ik, Ga{\v{s}}per},
  journal = {Neuron},
  volume  = {109},
  number  = {7},
  pages   = {1227--1241},
  year    = {2021}
}

@incollection{wagner2013robustness,
  title={Robustness and evolvability in living systems},
  author={Wagner, Andreas},
  booktitle={Robustness and evolvability in living systems},
  year={2013},
  publisher={Princeton university press}
}

@article{ciliberti2007robustness,
  title={Robustness can evolve gradually in complex regulatory gene networks with varying topology},
  author={Ciliberti, Stefano and Martin, Olivier C and Wagner, Andreas},
  journal={PLoS computational biology},
  volume={3},
  number={2},
  pages={e15},
  year={2007},
  publisher={Public Library of Science, San Francisco, USA}
}

@article{ma2009defining,
  title={Defining network topologies that can achieve biochemical adaptation},
  author={Ma, Wenzhe and Trusina, Ala and El-Samad, Hana and Lim, Wendell A and Tang, Chao},
  journal={Cell},
  volume={138},
  number={4},
  pages={760--773},
  year={2009},
  publisher={Elsevier}
}

\end{document}